\documentclass{aa}  
\usepackage{float}
\usepackage[varg]{txfonts}
\usepackage{amsmath}
\usepackage{graphicx}
\usepackage[table]{xcolor}
\usepackage{subcaption}
\usepackage{txfonts}
\usepackage{lipsum}
\usepackage{hyperref}
\hypersetup{
    colorlinks=true,
    linkcolor=black,
    filecolor=black,      
    urlcolor=black,
    citecolor=blue,
    }
\makeatletter
\renewcommand*\aa@pageof{, page \thepage{} of \pageref*{LastPage}}
\makeatother
\usepackage{orcidlink}

\begin{document} 

   \title{A normalizing flow approach for the inference of star cluster properties from unresolved broadband photometry}
   \titlerunning{A normalizing flow approach for the inference of star cluster properties}

   \subtitle{I. Comparison to spectral energy distribution fitting}

\author{
Daniel Walter\inst{1}\orcidlink{0009-0006-3668-9147}
\and Victor F.\ Ksoll\inst{1}\orcidlink{0000-0002-0294-799X}
\and Ralf S.\ Klessen\inst{1,2}\orcidlink{0000-0002-0560-3172}
\and Médéric Boquien\inst{3}\orcidlink{0000-0003-0946-6176}
\and Aida Wofford\inst{4}\orcidlink{0000-0001-8289-3428}
\and Francesco Belfiore\inst{5}\orcidlink{0000-0002-2545-5752}
\and Daniel A.\ Dale\inst{6}\orcidlink{0000-0002-5782-9093}
\and Kathryn Grasha\inst{7}\orcidlink{0000-0002-3247-5321}
\and David A.\ Thilker\inst{8}\orcidlink{0000-0002-8528-7340}
\and Leonardo \'Ubeda\inst{9}\orcidlink{0000-0001-7130-2880}
\and Thomas G.\ Williams\inst{10}\orcidlink{0000-0002-0012-2142}
}
        
\institute{Universität Heidelberg, Zentrum für Astronomie, Institut für Theoretische Astrophysik, Albert-Ueberle-Straße 2, 69120 Heidelberg, Germany
\and Universit\"{a}t Heidelberg, Interdisziplin\"{a}res Zentrum f\"{u}r Wissenschaftliches Rechnen, INF~225, 69120 Heidelberg, Germany
\and Université Côte d'Azur, Observatoire de la Côte d'Azur, CNRS, Laboratoire Lagrange, 06000, Nice, France
\and Instituto de Astronom\'ia, Universidad Nacional Aut\'onoma de M\'exico, Unidad Acad\'emica en Ensenada, Km 103 Carr. Tijuana--Ensenada, Ensenada, B.C.,
C.P. 22860, M\'exico
\and INAF -- Arcetri Astrophysical Observatory, Largo E. Fermi 5, I-50125, Florence, Italy
\and Department of Physics and Astronomy, University of Wyoming, Laramie, WY 82071, USA
\and Research School of Astronomy and Astrophysics, Australian National University, Canberra, ACT 2611, Australia
\and Department of Physics and Astronomy, The Johns Hopkins University, Baltimore, MD 21218, USA
\and Space Telescope Science Institute, Baltimore, MD, USA
\and Sub-department of Astrophysics, Department of Physics, University of Oxford, Keble Road, Oxford OX1 3RH, UK
  }

   \date{Received 1 August 2025; accepted 2 December 2025}

  \abstract
   {Estimating properties of star clusters from unresolved broadband photometry is a challenging problem that is classically tackled using spectral energy distribution (SED) fitting methods that are based on simple stellar population models. However, grid-based methods suffer from computational limitations. Because of their exponential scaling, they can become intractable when the number of inference parameters grows. In addition, nuisance parameters in the model can make the computation of the likelihood function intractable. These limitations can be overcome by modern generative deep learning methods that offer flexible and powerful tools for modeling high-dimensional posterior distributions and fast inference from learned data.}
   {We present a normalizing flow approach for the inference of cluster age, mass, and reddening parameters from \textit{Hubble} Space Telescope broadband photometry. In particular, we explore our network's behavior when dealing with an inference problem that has been analyzed in previous works.}
   {We used the SED modeling code \texttt{CIGALE} to create a dataset of synthetic photometric observations for $5 \times 10^6$ mock star clusters. Subsequently, this dataset was used to train a coupling-based flow in the form of a conditional invertible neural network to predict posterior probability distributions for cluster age, mass, and reddening from photometric observations.}
   {We predicted cluster parameters for the Physics at High Angular resolution in Nearby GalaxieS (PHANGS) Data Release 3 catalog. To evaluate the capabilities of the network, we compared our results to the publicly available PHANGS estimates and found that the estimates agree reasonably well.}
   {We demonstrate that normalizing flow methods can be a viable tool for the inference of cluster parameters, and argue that this approach is especially useful when nuisance parameters make the computation of the likelihood intractable and in scenarios that require efficient density estimation.}

   \keywords{Galaxies: star clusters: general --
                Methods: statistical --
                Methods: data analysis
               }

   \maketitle

\section{Introduction}\label{sec:introdution}
Star clusters form a link between the small-scale physics of star formation and stellar evolution and the large-scale physics of galactic environments and galaxy evolution. Young open clusters can inform us about the most recent star formation history of the host galaxy, while old globular clusters encode information about early galactic evolution. In light of this, there has been an effort to characterize the main parameters (i.e., age and mass) of galactic and extragalactic star clusters \citep[e.g.,][]{2005ApJ...631L.133F, Chandar_2016, 2019MNRAS.483.4949G, 2021MNRAS.502.1366T}.

High-resolution surveys of nearby galaxies ($d < 25 \:\mathrm{Mpc}$) that retrieve UV-to-optical broadband photometry, like the Legacy ExtraGalactic Ultraviolet Survey (LEGUS; \citealt{2015AJ....149...51C}) and Physics at High Angular resolution in Nearby GalaxieS (PHANGS) \textit{Hubble} Space Telescope (HST; \citealt{2022ApJS..258...10L, 2023ApJ...944L..17L}) surveys, enable the detection and classification of thousands of clusters per galaxy. However, at these distances, the stellar populations that constitute the clusters can typically not be resolved to the degree that is necessary for the construction of color-magnitude diagrams. Consequently, age and mass estimates are based on the integrated photometry of the clusters. By fitting the observed photometry to a grid of model photometry (under the assumption of an appropriate noise model), maximum-likelihood estimates (MLEs) of cluster parameters can be retrieved \citep[see, e.g.,][]{2021MNRAS.502.1366T}. Alternatively, the likelihood function can be multiplied by a suitable prior probability density to find the posterior distribution of cluster parameters (thus opting for Bayesian analysis). In the simplest case, the model photometry is calculated from a spectral energy distribution (SED) modeling code that assumes that the members of the cluster population have the same age and follow the initial mass function (IMF; i.e., a simple stellar population). In addition, dust attenuation can have a substantial effect on the UV and optical broadband fluxes and needs to be included to generate realistic SED models. In principle, there is a wide range of additional effects that can be included in the model (nebular emissions, stochastic IMF sampling for low-mass clusters, etc.) to create ever more realistic SEDs with the tradeoff of a higher computational cost. 

In this work we focus on the SED modeling code \texttt{CIGALE} \citep[Code Investigating GALaxy Emission;][]{2005MNRAS.360.1413B, 2009A&A...507.1793N, 2019A&A...622A.103B}, which is capable of producing mock star cluster photometry. Using a training set based on simulated \texttt{CIGALE} photometry, we attempted to model the Bayesian posterior probability distributions of the star cluster parameters with a deep learning approach based on the conditional invertible neural network (cINN) proposed by \citet{2019arXiv190702392A}. This specific network has already seen applications in image processing \citep{2019arXiv190702392A}, medicine \citep{10.1007/978-3-658-33198-6_80, jimaging7110243}, and astronomy \citep{10.1093/mnras/staa2931, 2022MNRAS.512..617K, refId0, Bister_2023, 10.1093/mnras/stad072, 2023A&A...674A.175K, 2023MNRAS.519.2199E, 2024A&A...683A.246K} and is based on the general framework of normalizing flows. We give a more detailed description of the method in Sect. \ref{sec:cINN recap}.

The \texttt{CIGALE} forward model is computationally inexpensive and can be used to directly compute the posterior probability distributions over a sufficiently dense grid via Bayes' theorem \citep{2021MNRAS.502.1366T}. For the simplest case, no neural network approach is needed to solve the cluster inference problem. However, the complexity of the forward model can drastically increase with the addition of nuisance parameters in the model. With nuisance parameters, we specifically mean quantities that we neither observe nor predict but whose stochasticity or uncertainty can nevertheless influence the result. For example, stochasticity in the distribution of stellar masses can have a large effect on the SED for low-mass clusters ($< 10^4 M_{\odot}$; \citealt{2012ApJ...745..145D, 2022MNRAS.509..522O}) and is consequently of crucial importance when estimating the properties of clusters with few stars. In general, there can be many such nuisance parameters in the model. Take, as an example, the $R_V$ parameter of the \citet{1989ApJ...345..245C} dust attenuation law. We might want to account for the fact that we do not know the true value of the $R_V$ parameter with certainty \citep[as suggested by the observed $R_V$ variability in the Milky Way;][]{2023ApJS..269....6Z}. A more honest model might account for the uncertainty by assuming that $R_V$ follows some distribution: $R_V \sim p(R_V)$ (deploying a probabilistic interpretation of $R_V$ in line with Bayesian reasoning). Since we do not observe $R_V$ directly and do not aim to predict it, it will effectively constitute a nuisance parameter of our model, similar to the stochasticity of the IMF. Of course, any other model parameter can, in principle, be treated in the same way.

This poses computational problems to the classic grid-based approaches that are ubiquitously used to retrieve cluster ages and masses. Formally, we can understand this by considering the likelihood function when we have a set of $n$ nuisance parameters $\boldsymbol{\theta}^* = (\theta_1^*, \dots, \theta_n^*)^T$:
\begin{equation}
    p(\mathbf{x} \mid \boldsymbol{\theta}) = \int p(\mathbf{x}, \boldsymbol{\theta}^* \mid \boldsymbol{\theta}) \: d\boldsymbol{\theta}^* =  \int p(\mathbf{x} \mid \boldsymbol{\theta}, \boldsymbol{\theta}^*) \cdot p(\boldsymbol{\theta}^*\mid \boldsymbol{\theta}) \: d\boldsymbol{\theta}^*,
\end{equation}
where $\boldsymbol{\theta}$ denotes the cluster parameters and $\mathbf{x}$ the photometric observations. Usually, the function $p(\mathbf{x} \mid \boldsymbol{\theta}, \boldsymbol{\theta}^*)$ is tractable but needs to be averaged over all combinations of $\boldsymbol{\theta}^* = (\theta_1^*, \dots, \theta_n^*)^T$. Adding a new grid dimension for every nuisance parameter quickly becomes intractable due to the ``curse of dimensionality.'' Alternatively, one can retrieve the likelihood as a Monte Carlo estimate:
\begin{equation}
    p(\mathbf{x} \mid \boldsymbol{\theta}) = \mathbb{E}_{\boldsymbol{\theta}^* \sim p(\boldsymbol{\theta}^*\mid \boldsymbol{\theta})} \left[ \: p(\mathbf{x} \mid \boldsymbol{\theta}, \boldsymbol{\theta}^*) \: \right] \approx \frac{1}{M}\sum_{i=1}^M p(\mathbf{x} \mid \boldsymbol{\theta}, \boldsymbol{\theta}_i^*).
\end{equation}
However, even this can be intractable if the variance of $p(\mathbf{x} \mid \boldsymbol{\theta}, \boldsymbol{\theta}^*)$ is sufficiently large, since the variance of the sample mean is given by $\mathrm{Var}_{\boldsymbol{\theta}^*} \left( p(\mathbf{x} \mid \boldsymbol{\theta}, \boldsymbol{\theta}^*) \right)/M$. In these cases, so-called likelihood-free approaches (sometimes referred to as simulation-based inference) are a natural choice since they do not rely on an explicit calculation of the likelihood \citep[see, e.g.,][]{2020PNAS..11730055C}. Instead, they require only the ability to generate output samples with the forward model. Normalizing flow approaches like the cINN can be trained in a likelihood-free manner and are therefore ideal candidates for these kinds of problems. In addition, the computational cost of normalizing flows is amortized. This means that an initial high computational cost (training, hyperparameter search, etc.) is compensated for by a low computational cost during inference. For example, in the context of galaxy parameter estimation, \citet{2022ApJ...938...11H} reduced the inference time per galaxy from 10--100 CPU hr to $\sim$ 1s. This property is especially useful when forward models become more complex (i.e., computationally expensive) and the number of inferences becomes large (e.g., when cluster catalogs become more extensive) and constitutes a particular advantage over the classical Markov chain Monte Carlo (MCMC) approach, which requires a re-simulation for every inference. We give a more detailed discussion of the computational advantages and disadvantages in Sect. \ref{sec:discussion}.

In recent years, normalizing flows have seen applications in a wide range of astronomical inference problems. Examples include stellar parameter estimation \citep{2023mla..confE..39Z, 2024A&A...692A.228C}, exoplanetary atmospheric retrieval \citep{2023A&A...672A.147V}, and the inference of galaxy properties \citep{2022MLS&T...3dLT04K}. Conceptually, our work is perhaps most similar to \citet{2022ApJ...938...11H}, who used masked autoregressive flows in conjunction with an SED modeling code to infer galaxy properties from broadband photometry.

To introduce normalizing flows for the inference of star cluster properties, we based our approach on a \texttt{CIGALE} model with a fully sampled IMF, a small number of photometric filters (five HST filters), and only three inference parameters (age, mass, and color excess ($E_{B-V}$), hereafter reddening). We wanted to explore the behavior of the cINN in this simplified setting before applying it to the more general and complex problem (i.e., models with nuisance parameters and a possibly larger number of photometric filters and inference parameters), which will be the focus of a follow-up paper. This allowed us to evaluate the performance of the network without additional confounding variables and enabled comparisons with previous works based on the standard approach of SED forward modeling and fitting. Specifically, we replicated the \texttt{CIGALE} setup of \citet{2021MNRAS.502.1366T} and compared our estimates to theirs. We want to stress that in this simplified setting, the normalizing flow approach is not necessary since posterior densities can be calculated directly over a relatively fine parameter grid.

We discuss the general normalizing flow approach, the cINN architecture, and the training procedure in Sect. \ref{sec:cINN recap}. Sections \ref{sec:synthetic_data} and \ref{sec:observed_data} describe the synthetic data (used for training) and the catalog of real cluster photometry \citep[used for the comparison with][]{2021MNRAS.502.1366T}. In Sect. \ref{sec:results} we present various tests and evaluations on a held-out test set from the synthetic dataset (Sect. \ref{sec:testing_on_synthetic_data}) and the real observations (Sect. \ref{sec:tests_on_phangs_catalog}). Finally, Sect. \ref{sec:discussion} provides a discussion of the advantages and disadvantages of the normalizing flow approach and compares it to other commonly used inference methods. The main results of our work are summarized in Sect. \ref{sec:summary}.

\section{Methods}\label{sec:methods}
\subsection{Normalizing flows and the cINN architecture}\label{sec:cINN recap}
Normalizing flows form a class of generative models that can be used to directly compute the probability density of the underlying dataset \citep{9089305}. This is in contrast to other methods like generative adversarial networks \citep[][]{NIPS2014_f033ed80} or variational autoencoders \citep[][]{2013arXiv1312.6114K}, which typically cannot be used to directly estimate densities. This is achieved by constructing an expressive mapping $f_{\boldsymbol{\phi}}$ that transforms a well-behaved base distribution (commonly a multivariate standard normal) into the target distribution (i.e., the distribution of the training set): $f_{\boldsymbol{\phi}} (\mathbf{Z}) \approx \boldsymbol{\Theta}$, where $\boldsymbol{\Theta}$ denotes the target variables and $\boldsymbol{\phi}$ the collection of free parameters of $f$ (e.g., the weights and biases of a neural network). In our case, $\boldsymbol{\Theta}$ refers to the cluster parameters for which we want to model the posterior probability density. The variables $\mathbf{Z} \sim \mathcal{N}(0, \mathbb{I})$ are commonly called ``latent variables.'' Usually, $\boldsymbol{Z}$ does not have a direct physical meaning. Under the condition that $f_{\boldsymbol{\phi}}$ is invertible and differentiable, the density of $f_{\boldsymbol{\phi}} (\mathbf{Z})$ can be calculated from the density of the latent variables via the change-of-variables formula:
\begin{equation}\label{eq:flow_density}
    p_{f_{\boldsymbol{\phi}}(\mathbf{Z})}(\boldsymbol{\theta}) = \frac{p_{\mathbf{Z}}\left(f_{\boldsymbol{\phi}}^{-1}(\boldsymbol{\theta})\right)}{\left|\det J_{f_{\boldsymbol{\phi}}}\left(f_{\boldsymbol{\phi}}^{-1}(\boldsymbol{\theta})\right)\right|},
\end{equation}
where $J_{f_{\boldsymbol{\phi}}}$ denotes the Jacobi matrix of $f_{\boldsymbol{\phi}}$. The network parameters $\boldsymbol{\phi}$ can then be retrieved by maximum-likelihood estimation --- or equivalently by minimizing the Kullback–Leibler divergence $D_{KL}(p_{\boldsymbol{\Theta}} \parallel p_{f_{\boldsymbol{\phi}}(\mathbf{Z})})$ (see, e.g., \citealt{JMLR:v22:19-1028}) --- which leads to a negative log-likelihood function of the form
\begin{equation}
    -\log L(\boldsymbol{\phi}) \propto \: \frac{1}{N}\sum_{i=1}^N \left(\frac{ ||f_{\boldsymbol{\phi}}^{-1}(\boldsymbol{\theta}_i) ||^2}{2} + \log \left|\det J_{f_{\boldsymbol{\phi}}}\left(f_{\boldsymbol{\phi}}^{-1}(\boldsymbol{\theta}_i)\right) \right| \right),
\end{equation}
where $L(\boldsymbol{\phi})$ denotes the likelihood function relative to some training set $\{\boldsymbol{\theta}_1, \dots, \boldsymbol{\theta}_N \}$.

Once the network is trained, it can be used for both density estimation (using Eq.~\ref{eq:flow_density}) and sampling. To generate samples of the target distribution, one simply draws samples from the latent variable distribution $\mathbf{z}_1, \dots, \mathbf{z}_k \sim \mathcal{N}(0, \mathbb{I})$ and transforms them to the target space $f_{\boldsymbol{\phi}}(\mathbf{z}_1), \dots, f_{\boldsymbol{\phi}}(\mathbf{z}_k)$. Note that sampling requires the evaluation of the forward direction of the network, while density estimation requires the backward direction.

In practice, one is usually not interested in modeling a single probability density distribution but an entire family of conditional probability density distributions. In that case, $f_{\boldsymbol{\phi}}$ will have an additional input $\mathbf{x}$ (i.e., $f_{\boldsymbol{\phi}}(\boldsymbol{\theta}; \mathbf{x})$) that denotes the variables over which the probability distribution is conditioned (in our case the cluster photometry).

\begin{figure}
    \centering
    \includegraphics[trim={1.1cm 0 1.1cm 0}, clip, width=\columnwidth]{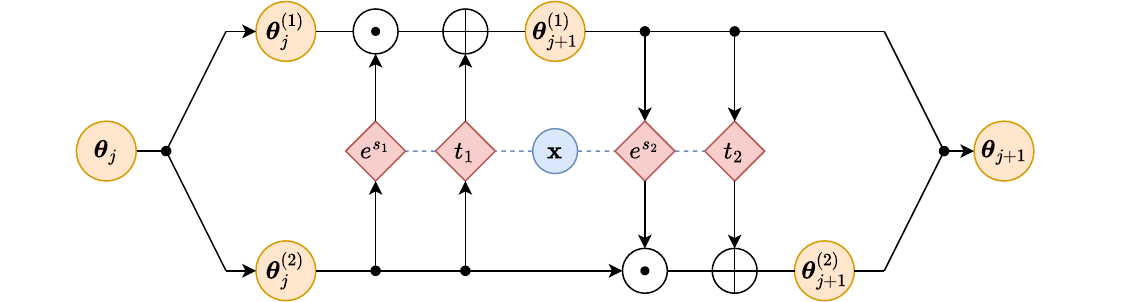}
    \caption{Flowchart of single coupling block (representing Eqs. \ref{eq:coupling_equation_1}, \ref{eq:coupling_equation_2}, and \ref{eq:coupling_equation_3}) taken from \citet{MasterThesis}. The input vector, $\boldsymbol{\theta}_j$, is split into two sub-vectors, $\boldsymbol{\theta_j}^{(1)}$ and $\boldsymbol{\theta}_j^{(2)}$. The conditioning input, $\mathbf{x,}$ is used as an additional input to the subnetworks $s_1$, $s_2$, $t_1,$ and $t_2$.}
    \label{fig:coupling_block}
\end{figure}

For the mapping $f_{\boldsymbol{\phi}}$, we used the cINN architecture of \citet{2019arXiv190702392A}, which is based on the composition of invertible coupling blocks that have a tractable determinant of the Jacobian. For an input vector $\boldsymbol{\theta}_j$ (in the $j$-th network layer), the coupling blocks split the vector into two sub-vectors $\boldsymbol{\theta}_j = \{\boldsymbol{\theta}_j^{(1)}, \boldsymbol{\theta}_j^{(2)}\}$ and perform the transformation
\begin{flalign}
        &\boldsymbol{\theta}_{j+1}^{(1)} = \boldsymbol{\theta}_j^{(1)} \odot\: \exp \left(s_1\left(\boldsymbol{\theta}_j^{(2)}, \mathbf{x}\right)\right) \oplus t_1\left(\boldsymbol{\theta}_j^{(2)}, \mathbf{x}\right)&\label{eq:coupling_equation_1}\\
        &\boldsymbol{\theta}_{j+1}^{(2)} = \boldsymbol{\theta}_j^{(2)} \odot\: \exp \left(s_2\left(\boldsymbol{\theta}_{j+1}^{(1)}, \mathbf{x}\right)\right) \oplus t_2\left(\boldsymbol{\theta}_{j+1}^{(1)}, \mathbf{x}\right)&\label{eq:coupling_equation_2}\\
        &\boldsymbol{\theta}_{j+1} = \left\{\boldsymbol{\theta}_{j+1}^{(1)},\: \boldsymbol{\theta}_{j+1}^{(2)}\right\},\label{eq:coupling_equation_3}&
\end{flalign}
where $\odot$ and $\oplus$ denote the component-wise multiplication and addition. The mappings $s_1$, $s_2$, $t_1$, and $t_2$ are realized by fully connected multilayer neural networks. We can visualize this mapping in a flowchart (see Fig. \ref{fig:coupling_block}), from which it is apparent that the coupling block transformation is indeed invertible. For further details on the expressibility and the Jacobian matrix of coupling-based architectures, we refer to \citet{dinh2017density}.
In actuality, the four subnetworks --- $s_1$, $s_2$, $t_1$, and $t_2$ --- are combined to two subnetworks ---$s_1$/$t_1$ and $s_2$/$t_2$ --- that produce both the multiplicative and additive output inside the coupling block. With the successive composition of these coupling blocks, we can create highly expressive mappings. To further increase the expressibility, the output vector components are permuted before they are passed to the next coupling block. These permutations are randomly initialized but fixed during training and testing, i.e., there is the same permutation for every run. 

To minimize the negative log-likelihood, $-\log L(\phi)$, during training, we used the gradient descent based Adam optimizer \citep{2014arXiv1412.6980K} and an additional regularization term $\eta \cdot ||\boldsymbol{\phi}||_2^2$ to prevent overfitting. To optimize the hyperparameters of the network, we used the \texttt{HYPERBAND} algorithm, as proposed by \citet{JMLR:v18:16-558}. A full list of hyperparameters is given in Appendix \ref{network_parameters}. The network is implemented through the PyTorch-based \citep{NEURIPS2019_bdbca288} Framework for easily Invertible Architectures \citep[FrEIA;][]{freia}.

\subsection{Synthetic data}\label{sec:synthetic_data}
To generate the synthetic training set, we used the PHANGS branch of the \texttt{CIGALE} GitLab repository\footnote{\url{https://gitlab.lam.fr/cigale/cigale/-/tree/PHANGS?ref_type=heads}; git commit hash: 9dd1827ae3fc67a1d412153309d924650abf110f; last accessed: May 2025} and reproduced the configuration setup of \citet{2021MNRAS.502.1366T}. In this setup, the simple stellar population  spectra are based on the \citet{2003MNRAS.344.1000B} model and the star formation history is modeled as an instantaneous burst. To account for dust, the SED is reddened by a Milky Way extinction screen using the \citet{1989ApJ...345..245C} law. Nebular emissions of young clusters are not accounted for.

For the synthetic photometry, we considered the five HST WFC3-UVIS filters used in the PHANGS survey, namely F275W (near-UV), F336W (U), F438W (B), F555W (V), and F814W (I), ranging from the UV to the near-infrared. In this wavelength range, dust emission is practically absent and therefore neglected.

We assumed a prior distribution of cluster parameters that is log-uniform for age ($t$) and initial mass ($m_0$), and uniform for reddening ($E_{B-V}$), i.e., $\log_{10} t \sim \mathcal{U}$, $\log_{10} m_0 \sim \mathcal{U}$ and $E_{B-V} \sim \mathcal{U}$. The parameters are assumed to be independent in the prior with parameter ranges $1 \:\mathrm{Myr} < t < 13750 \:\mathrm{Myr}$, $10^2 \:\mathrm{M_{\odot}} < m_0 < 10^8 \:\mathrm{M_{\odot}}$ and $0 \:\mathrm{mag} <E_{B-V} < 1.5 \:\mathrm{mag}$.  This implies that the prior is jointly uniform over the parameters $\log_{10} t$, $\log_{10} m_0$ and $E_{B-V}$. We note that this distribution is not physically motivated. Instead, it has the advantage of allowing us to easily compare our maximum a posteriori (MAP) estimates with the PHANGS MLEs. Under the assumption of a uniform prior, the two estimates must be equivalent:
\begin{equation}\label{eq:equivalence_map_ml}
    p(\boldsymbol{\theta}) = \mathcal{U}(\boldsymbol{\theta}) \implies p(\boldsymbol{\theta}\:|\: \mathbf{x}) \propto p(\mathbf{x}\:|\:\boldsymbol{\theta}) \implies \mathrm{MAP} = \mathrm{MLE}.
\end{equation}
Note also that these priors do not replicate the priors that \citet{2021MNRAS.502.1366T} use in their Bayesian analysis. Therefore, we only replicated their MLEs and not their Bayesian estimates.

From this distribution, we sampled $\approx 5 \times 10^6$ mock star clusters, for which we simulated broadband photometry with \texttt{CIGALE}. From these, $\approx 10^5$ example clusters are held out as a test set and an additional $\approx 10^5$ are held out as a validation set for the hyperparameter search.

Our \texttt{CIGALE} setup is limited to an age resolution of 1 Myr. We find that the network learns to reproduce this grid structure, which results in undesirable peaks in the posterior at the positions of the grid points. This also leads to computational problems for the flow model, since the probability density at these points is essentially infinite. To alleviate this, we added a uniform noise to the age grid so that the probability mass at these points is distributed over the entire valid range: $t' = t + \epsilon$, with $\epsilon \sim \mathcal{U}_{[-0.5, 0.5]}$. This prevents the network from overfitting on the parameter grid.

By default, \texttt{CIGALE} generates photometry normalized to a reference distance of $d_0 = 10 \:\mathrm{pc}$ (for redshift $z=0$). We wanted the network to be able to generate parameter estimates for a range of distances. Thus, for every photometric value in the synthetic dataset, we randomly sampled a distance $d \sim \mathcal{U}_{[1 \: \mathrm{Mpc},\: 30 \: \mathrm{Mpc}]}$ and corrected the photometric fluxes via $f' = f \cdot \left(d_0/d\right)^2$. The parameter $d$ is then cast as an additional conditioning input.

\begin{figure*}
    \centering
    \includegraphics[width=\textwidth]{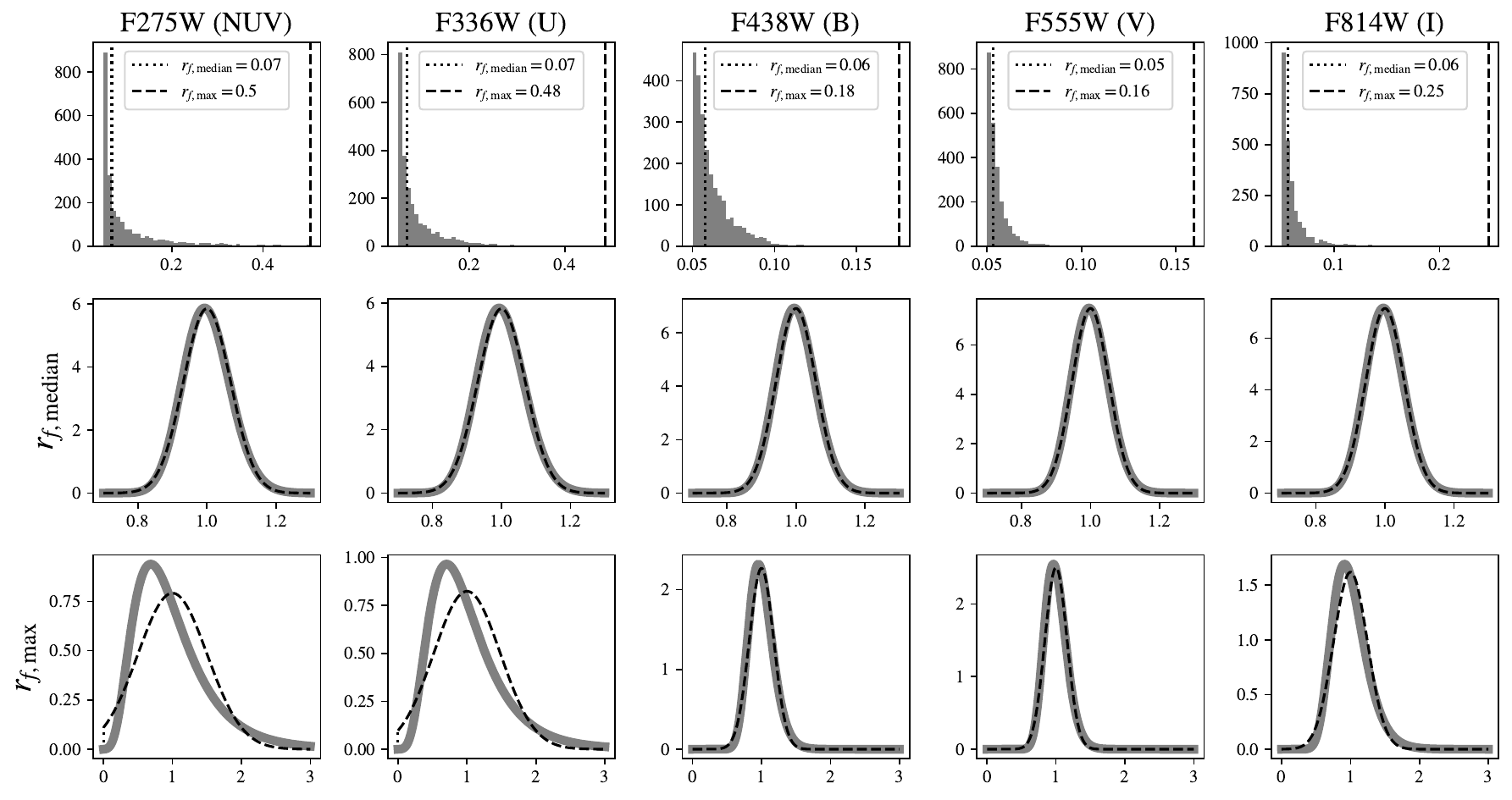}
    \caption{Top row: Histograms of relative flux uncertainties in the PHANGS cluster catalog. The dotted vertical lines denote the median relative flux uncertainties, and the dashed lines denote the maximum relative flux uncertainties. Middle row: Comparison of the normal noise model and the log-normal noise model for median relative flux uncertainties. Bottom row: Comparison of the normal noise model and the log-normal noise model for maximum relative flux uncertainties.}
    \label{results:fig:flux_uncertainty_statistics}
\end{figure*}

To account for measurement uncertainties, we added random noise to the photometry: for every mock cluster photometry $f,$ we sampled a relative error $r \sim \mathcal{U}_{[0, \:0.5]}$ (i.e., a maximum relative error of 50\%), from which we retrieved an uncertainty $\sigma = f \cdot r$. We then generated a noisy observation by sampling from a log-normal distribution with mean $f$ and standard deviation $\sigma$, i.e., $f^* \sim \mathcal{L}\mathcal{N}(f, \sigma)$ (this is equivalent to a normally distributed noise in magnitude space). The observed relative error $r^* = \sigma/f^*$ is saved and passed to the network as an additional conditioning input. That way, the cINN learns how a wide range of measurement uncertainties affect the photometry. Note that we deviated from \citet{2021MNRAS.502.1366T} here, who assumed a normally distributed noise on the fluxes. This is mainly motivated by a technical issue that we encounter when training the network. We find the best results when the fluxes are logarithmically scaled before passing them to the network. Naturally, this is only possible if the fluxes are strictly positive. A log-normal noise model guarantees exactly that. Normally distributed noise will generally lead to negative flux values in the training set. Of course, it is possible to shift the resulting distribution to positive values and apply a log-transformation afterward. However, we found that the resulting preprocessing led to poorer overall performance. Using a positively truncated normal distribution is also possible, but results in a distribution with a systematically lower standard deviation. This can be corrected by rescaling the distribution, but again leads to a noise model that is not equivalent to the approach of \citet{2021MNRAS.502.1366T}. Finally, we settled on the log-normal noise model, since it provided the easiest treatment of the flux uncertainties. Figure \ref{results:fig:flux_uncertainty_statistics} shows the distribution of relative flux uncertainties in the cluster catalog for the five photometric bands, as well as the resulting noise models for the median and maximum relative flux uncertainties. At median relative flux uncertainty, the log-normal noise model is practically indistinguishable from the normal noise model. Only for the highest relative uncertainties can we see substantial deviations in the near-UV and U band. As we will see, this can lead to considerable deviations in the parameter estimates for the clusters with the highest relative flux uncertainties. For the vast majority of clusters, however, the log-normal noise model is effectively equivalent to the normal noise model.

Since the generated fluxes span many orders of magnitudes, it is useful to represent them in log-space before passing them to the network. Overall, this results in three predicted parameters ($\log_{10} t$, $\log_{10} m$ and $E_{B-V}$) and eleven conditioning inputs ($\log f^*_\mathrm{F275W}$, $\log f^*_\mathrm{F336W}$, $\log f^*_\mathrm{F438W}$ $\log f^*_\mathrm{F555W}$, $\log f^*_\mathrm{F814W}$, $r^*_\mathrm{F275W}$, $r^*_\mathrm{F336W}$, $r^*_\mathrm{F438W}$, $r^*_\mathrm{F555W}$, $r^*_\mathrm{F814W}$, and $d$).

A technical difficulty arises from the fact that the prior distribution enforces sharp bounds on the parameter values. We found that it is difficult for the cINN to transform the standard normal distribution into a sharply bounded posterior distribution, which resulted in the network generating values outside the boundaries of the training set (e.g., negative $E_{B-V}$ values). This problem can be solved by an additional data preprocessing step that bijectively maps a bounded support onto $\mathbb{R}$ (consequently the inverse will always be within the bounds). If $\Theta$ is a uniformly distributed random variable with support $[a, b]$, i.e., $\Theta \sim \mathcal{U}_{[a,b]}$, then we have $(\Theta-a)/(b-a) \sim \mathcal{U}_{[0, 1]}$ and consequently
\begin{equation}
    \Phi^{-1}\left(\frac{\Theta-a}{b-a}\right) \sim \mathcal{N}(0, 1),
\end{equation}
where $\Phi(u) = \frac{1}{2} \cdot \left( 1 + \mathrm{erf}\left(u/\sqrt{2}\right) \right)$ denotes the cumulative distribution function of the standard normal distribution. We used this invertible mapping component-wise on $\log_{10} t$, $\log_{10} m$, and $E_{B-V}$ to enforce the strict bounds of the prior.

The generation of the synthetic dataset and training of the neural network were performed on the Compute Servers of the Interdisciplinary Center for Scientific Computing\footnote{\url{https://typo.iwr.uni-heidelberg.de/services/facilities/compute-server.html} (last accessed: Nov 2025)} (IWR) in Heidelberg. For the \texttt{CIGALE} simulation runs, we used four 12-Core AMD Opteron 6176SE 2.3 GHz processors ($\approx 11 \:\mathrm{h}$) and for the network training two 2 x 14-Core Intel Xeon Gold 6132 processors in conjunction with an Nvidia Titan Xp graphics card, taking $\approx 1.5 - 3 \:\mathrm{h}$ for a single full training run (depending on the size of the network).

\subsection{Observed data}\label{sec:observed_data}
The observed photometry is taken from the PHANGS Data Release 3 catalog \citep[Catalog Release 1;][]{2022ApJS..258...10L, 2022MNRAS.509.4094T}, which are publicly available at the Mikulski Archive for Space Telescopes (MAST)\footnote{\url{https://archive.stsci.edu/hlsp/phangs/phangs-cat} (last accessed: Nov 2025)}. These catalogs contain photometry for clusters in five galaxies (NGC 1433, NGC 1559, NGC 1566, NGC 3351, and NGC 3627) as well as MLEs for age, mass, and reddening based on \texttt{CIGALE}s fitting procedure. Note that by now, a cluster catalog over the full PHANGS-HST galaxy sample (38 galaxies) is available \citep{2024ApJS..273...14M}.

PHANGS provides catalogs of both human- and machine-classified clusters that are distinguished into Class 1 (symmetric, centrally concentrated, radial profile extended relative to point source) and Class 2 (asymmetric, centrally concentrated, radial profile extended relative to point source) type clusters. We included both Class 1 and Class 2 type clusters in our analysis but limited ourselves to the human-classified catalogs, which contain 2638 clusters in total. To simplify the evaluation with the cINN, we only kept those clusters that have information in all five photometric bands, reducing the total dataset size by $7.5 \%$ to 2439. In order to replicate the approach of \citet{2021MNRAS.502.1366T}, the galaxy distances were taken from LEGUS \citep[][]{2015AJ....149...51C} and uncertainties in the distances were neglected. Note, however, that better distance estimates are available \citep{2021MNRAS.501.3621A}. Finally, to fully replicate the approach of \citet{2021MNRAS.502.1366T}, we added an additional $5 \%$ systematic error to the flux uncertainties:
\begin{equation}
    {r^{*}}^{\prime} = \sqrt{{r^{*}}^2 + 0.05^2}.
\end{equation}

\section{Results}\label{sec:results}
\subsection{Tests on synthetic data}\label{sec:testing_on_synthetic_data}
Before we applied the network to the real-world photometry of the PHANGS catalog, we investigated its behavior on the synthetic test set. In contrast to the real-world photometry, the synthetic dataset contains the ground truth values for the cluster parameter. This allowed us to perform calibration checks on the synthetic data that we cannot perform on the PHANGS catalog (for which ground truth parameter values are unavailable). A popular diagnostic tool for assessing the quality of Bayesian inference algorithms is simulation-based calibration \citep[SBC;][]{2018arXiv180406788T}. The SBC method proceeds as follows: consider a set of simulated test parameters and observations $\{(\boldsymbol{\theta}_1, \mathbf{x}_1), \dots, (\boldsymbol{\theta}_m, \mathbf{x}_m)\}$ (i.e., a synthetic dataset that is not part of the training set of the network).  For every test observation $\mathbf{x}_i$, generate a number of posterior samples using the generative model: $\boldsymbol{\theta}_{i,1}', \dots, \boldsymbol{\theta}_{i,n}' \sim p_{\mathrm{cINN}}(\boldsymbol{\theta} \mid \mathbf{x}_i)$. Now let $T(\boldsymbol{\theta})$ be any statistic of the parameters ($\boldsymbol{\theta}$). \citet{2018arXiv180406788T} show that the rank of $T(\boldsymbol{\theta_i})$ in $\{T(\boldsymbol{\theta}_{i,1}'), \dots, T(\boldsymbol{\theta}_{i,n}')\}$ must be uniformly distributed if the generative model reproduces the correct posterior distributions (i.e., $p_{\mathrm{cINN}}(\boldsymbol{\theta} \mid \mathbf{x}) = p(\boldsymbol{\theta} \mid \mathbf{x})$). Deviations from uniformity are therefore an indication of modeling errors. Note that the reverse implication is generally not true (i.e., one may have completely wrong posteriors that nevertheless produce a perfectly uniform distribution over the ranks of some statistic $T$). We performed SBC with two sets of statistics. The first set is given by $T_1(\boldsymbol{\theta}) = t$, $T_2(\boldsymbol{\theta}) = m$ and $T_3(\boldsymbol{\theta}) = E_{B-V}$. Deviations from uniformity for these statistics indicate issues with the marginal posterior distributions. The second set of statistics will characterize the relationships between pairs of parameters: $T_4(\boldsymbol{\theta}) = \left(\log t \right) \cdot \left( \log m \right)$, $T_5(\boldsymbol{\theta}) =\left( \log m \right) \cdot E_{B-V}$ and $T_6(\boldsymbol{\theta}) = \left( \log t \right) \cdot E_{B-V}$. Here, deviations from uniformity may indicate problems in the network's modeling of the posterior covariances. The two sets of statistics are therefore complementary.

\begin{figure}[!ht]
    \centering
    \includegraphics[width=0.5\textwidth]{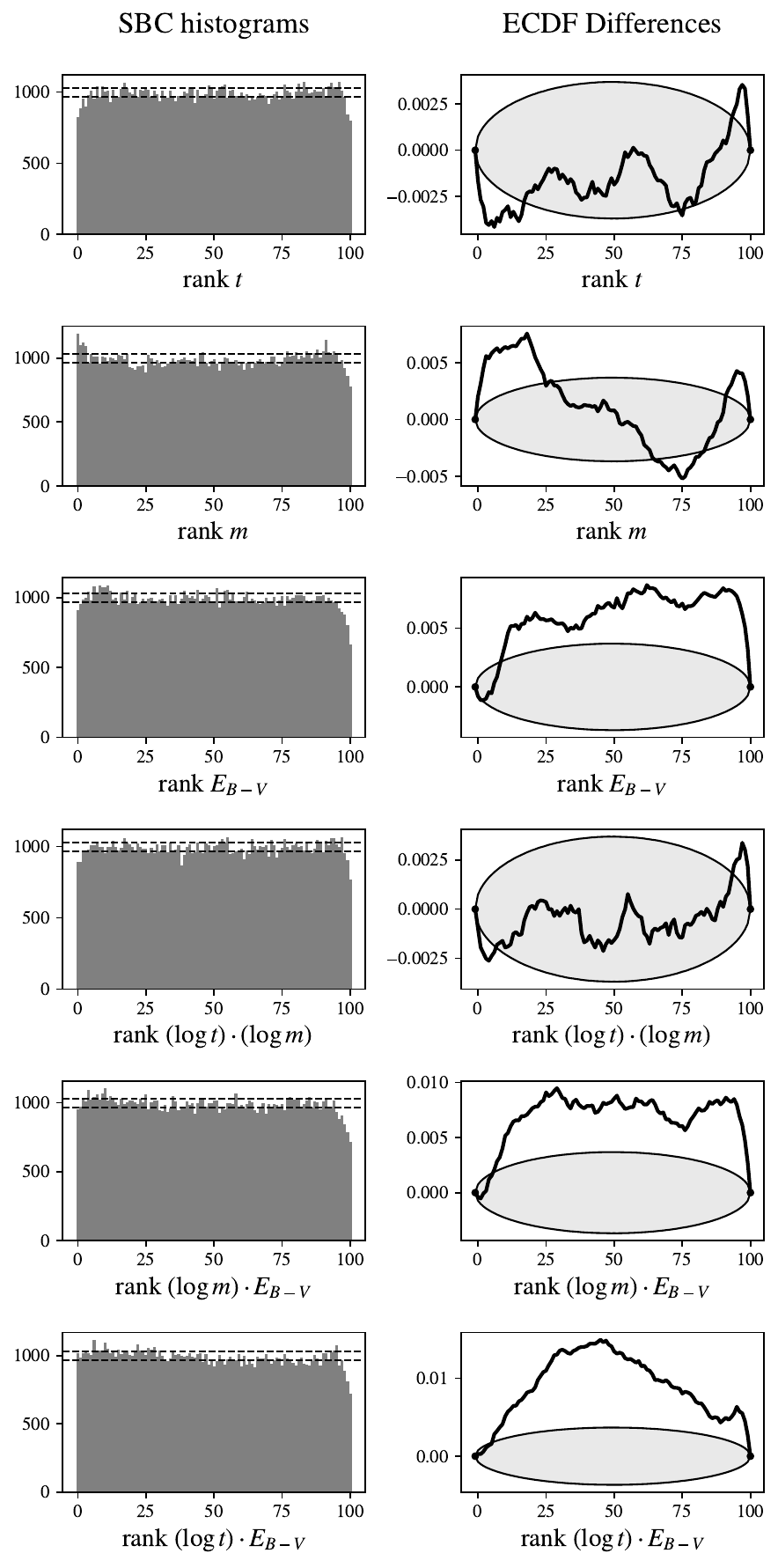}
    \caption{Results of SBC analysis over the entire test set ($\approx 10^5$ parameter--observation pairs). We generated a hundred posterior samples for every element in the test set. Left column: Rank histograms for the proposed statistics. Horizontal dashed lines correspond to the $\pm1\sigma$ range under the assumption of uniformity. Right column: Difference between rank ECDF and uniform cumulative distribution function. The gray area highlights the expected 0.01-0.99 quantile range of the ECDF difference under the assumption of uniformity.}
    \label{fig:sbc}
\end{figure}

Figure \ref{fig:sbc} (left column) shows the corresponding SBC rank histograms.
As we can see, the rank distributions are roughly uniform. To emphasize deviations at small and large ranks, \citet{2018arXiv180406788T} suggest pairing the rank histograms with plots of the empirical cumulative distribution function (ECDF). In Fig. \ref{fig:sbc} (right column), we show the differences between the rank ECDF and the uniform cumulative distribution function together with the 0.01-0.99 quantile range (gray area) that we would expect under perfect uniformity (assuming a binomial distribution for the values of the ECDF). The fact that the curves lie (partially) outside of the gray area, suggests a nonuniformity in the rank distribution. We can use the shape of the curves to get a rough idea of the type of failure. For example, the statistic $T_2$ shows a slight over-concentration at low and high ranks (except for very high ranks, where we have an under-concentration) indicating that the marginal mass distributions are, on average, slightly too narrow. The statistic $T_3$ shows an ECDF difference plot that is mostly above the expected range. This suggests that the network is, on average, overestimating the cluster reddening. The same pattern is seen in the ECDF difference plots for $T_5$ and $T_6$. This may indicate that the network is generating covariances between mass and reddening, as well as age and reddening, that are somewhat larger than the covariances in the test set. However, an alternative interpretation is simply that the overestimates in $E_{B-V}$ produce somewhat larger values in the product statistics. This is supported by the fact that the overestimation does not occur for the statistic $T_4$ (which does not use the reddening).

\subsection{Tests on PHANGS catalog}\label{sec:tests_on_phangs_catalog}
\subsubsection{Posterior predictive checks}
Unlike the synthetic data, the real-world photometry of the PHANGS catalog does not provide us with ground truth parameter values. This means that the SBC approach of Sect. \ref{sec:testing_on_synthetic_data} cannot be applied to the dataset of real-world photometry. In this case, a common approach is given by posterior predictive checks \citep[PPCs;][]{90a5941c-cead-3092-a18f-09867c5c72fb, 62bfc978-09b1-3997-9776-380d0b45e9c2}: Suppose $\{\mathbf{x}_1, \dots, \mathbf{x}_m\}$ is a set of observations. For every observation $\mathbf{x}_i$, we generated samples of the posterior from the generative model: $\boldsymbol{\theta}_{i,1}', \dots, \boldsymbol{\theta}_{i,n}' \sim p_{\mathrm{cINN}}(\boldsymbol{\theta} \mid \mathbf{x}_i)$. Then we re-simulated these parameters to generate corresponding synthetic observations: $\mathbf{x}_{i,j}'  \sim p(\mathbf{x} \mid \boldsymbol{\theta}_{i,j})$. The resulting distribution is commonly referred to as the posterior predictive distribution of $\mathbf{x}_i$. We could then compare the posterior predictive distribution with the initial observation. On average, the posterior predictive distribution should not deviate strongly from the true observation.
\begin{figure}[!ht]
    \centering
    \includegraphics[width=0.5\textwidth]{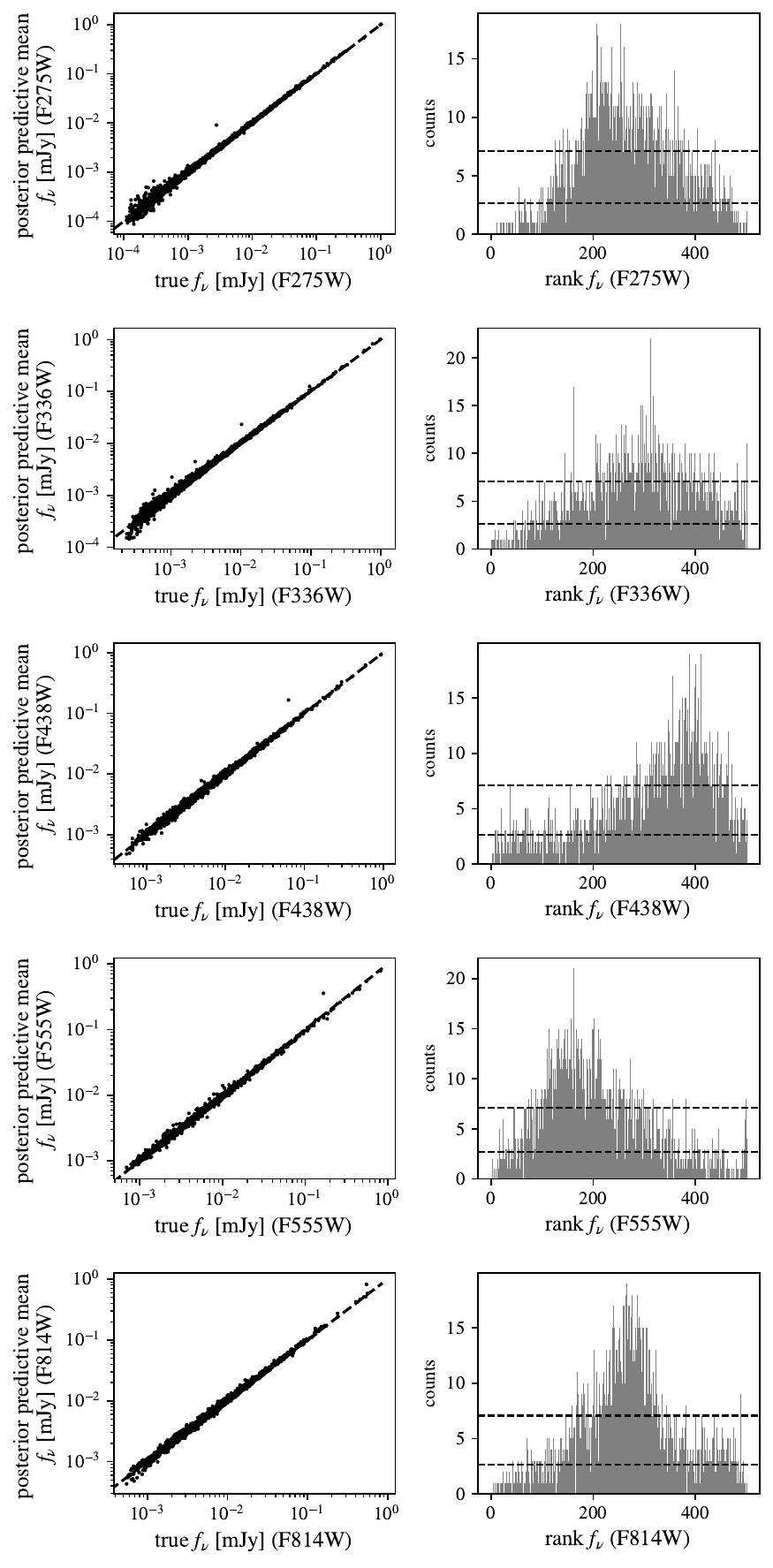}
    \caption{Results of PPC analysis. Posterior predictive distributions with 500 samples each were generated for all clusters in the PHANGS catalog. Left column: Comparison of the mean predictive photometry and the initial photometry. Right column: Rank distribution of the initial photometry within the posterior predictive distribution. Horizontal dashed lines correspond to the $\pm1\sigma$ range under the assumption of uniformity.}
    \label{fig:ppc}
\end{figure}
Figure \ref{fig:ppc} (left column) shows a comparison between the mean posterior predictive photometry and the initial PHANGS photometry for the five photometric bands. As we can see, the mean of the re-simulated photometry tends to be close to the ground truth. In Fig. \ref{fig:ppc} (right column) we show the rank of the PHANGS photometry within the posterior predictive distribution. Note that high and low ranks tend to be underrepresented, while mid-range ranks tend to be overrepresented. This suggests that the PHANGS photometry tends to lie well within the posterior predictive distribution. Note that we do not expect a uniform distribution of ranks in our PPC analysis. In fact, p-values of posterior predictive distributions are known to be ``stochastically less variable,'' i.e., more concentrated toward mid-range ranks \citep[see Sect. 6.3 of][]{gelman2013bayesian}.

\subsubsection{Example cases}
The absence of nuisance parameters in the model makes the computation of the likelihood tractable. This allows us to compute the posteriors directly. In this section we compare the cINN generated posteriors with the ground truth posteriors that we get from direct computation via Bayes' theorem. To that end, we chose four example clusters that we used for a density comparison and that are presented in Fig.~\ref{results:fig:posterior_examples_2d}. We chose these examples to cover a wide cluster age range. In addition, we highlight the cINN MAP estimates (triangle) and the PHANGS MLEs (cross) in the plots.
\begin{figure*}
    \includegraphics[width=\textwidth]{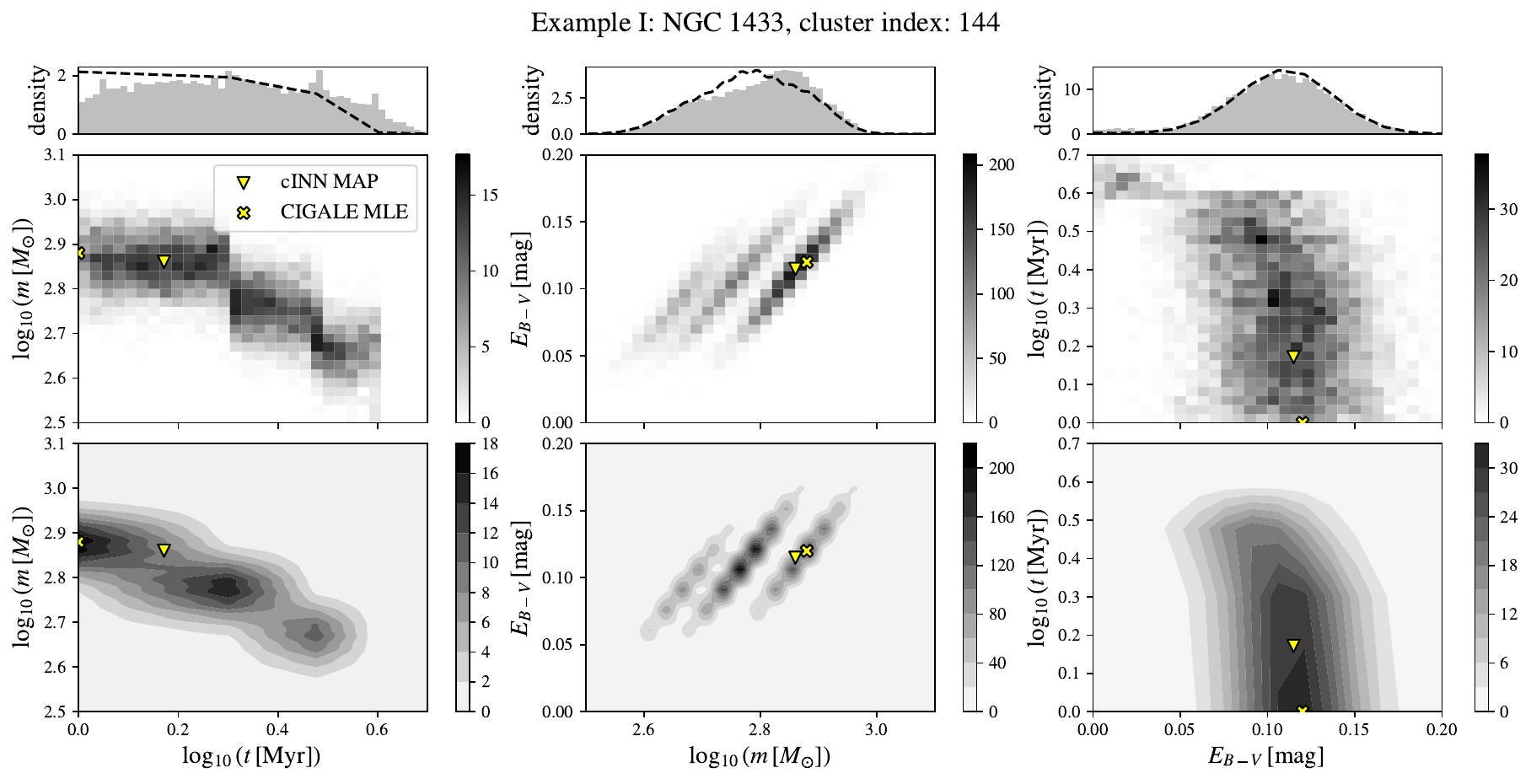}
    \includegraphics[width=\textwidth]{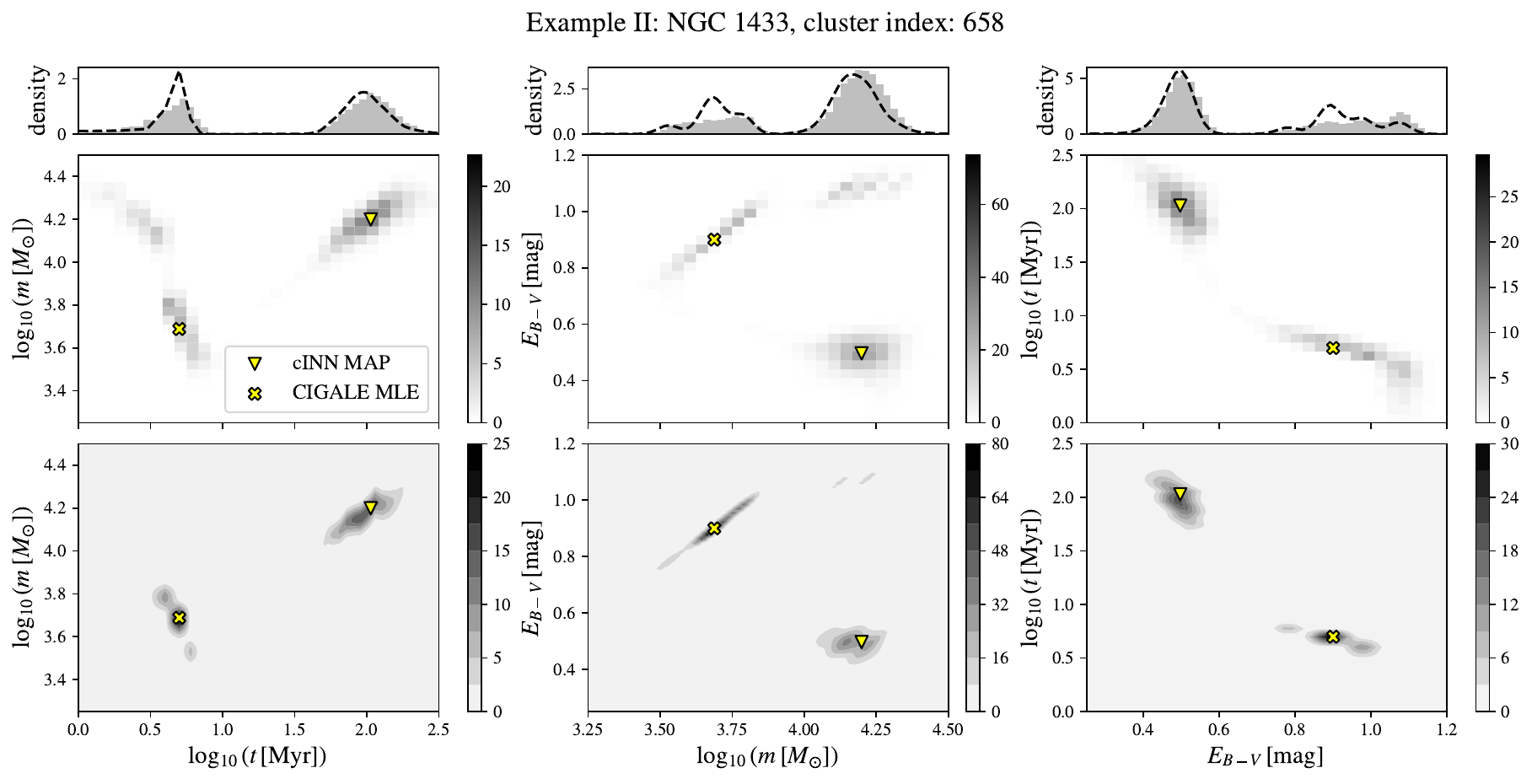}
    \caption{2D and 1D density plots of example posteriors. Top rows: Marginal 1D density for the cINN posterior (gray histogram) and ground truth (dashed black line). Middle rows: 2D density of the cINN generated posterior ($10^4$ samples per posterior). Bottom rows: Contour plot of ground truth densities calculated over a parameter grid. Note that we use the same color maps in the second and third rows (although the colorbars show different ranges). The yellow markers indicate positions of cINN MAP estimates (triangle) and PHANGS MLEs (cross).}
    \label{results:fig:posterior_examples_2d}
\end{figure*}
\begin{figure*}
    \ContinuedFloat
    \includegraphics[width=\textwidth]{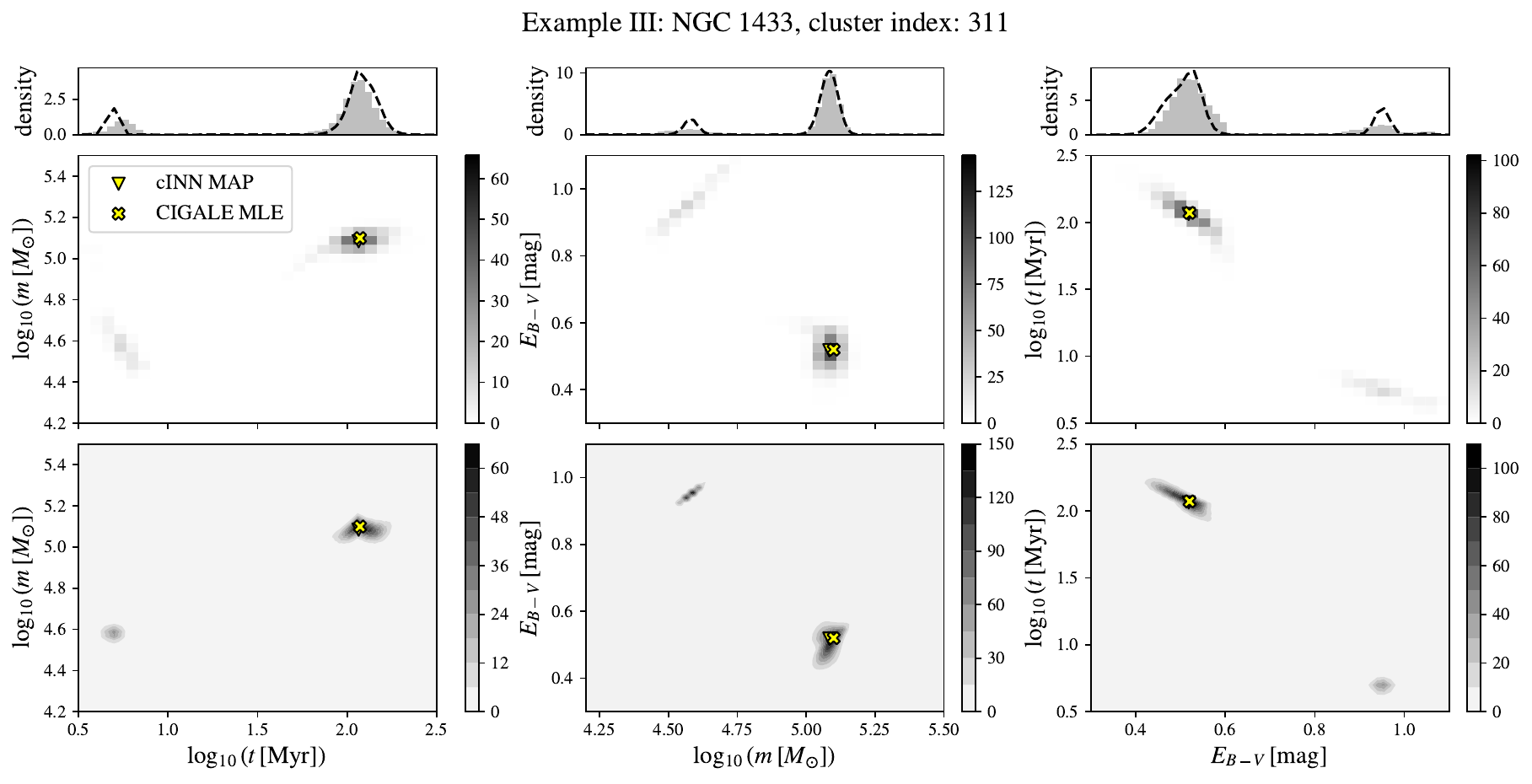}
    \includegraphics[width=\textwidth]{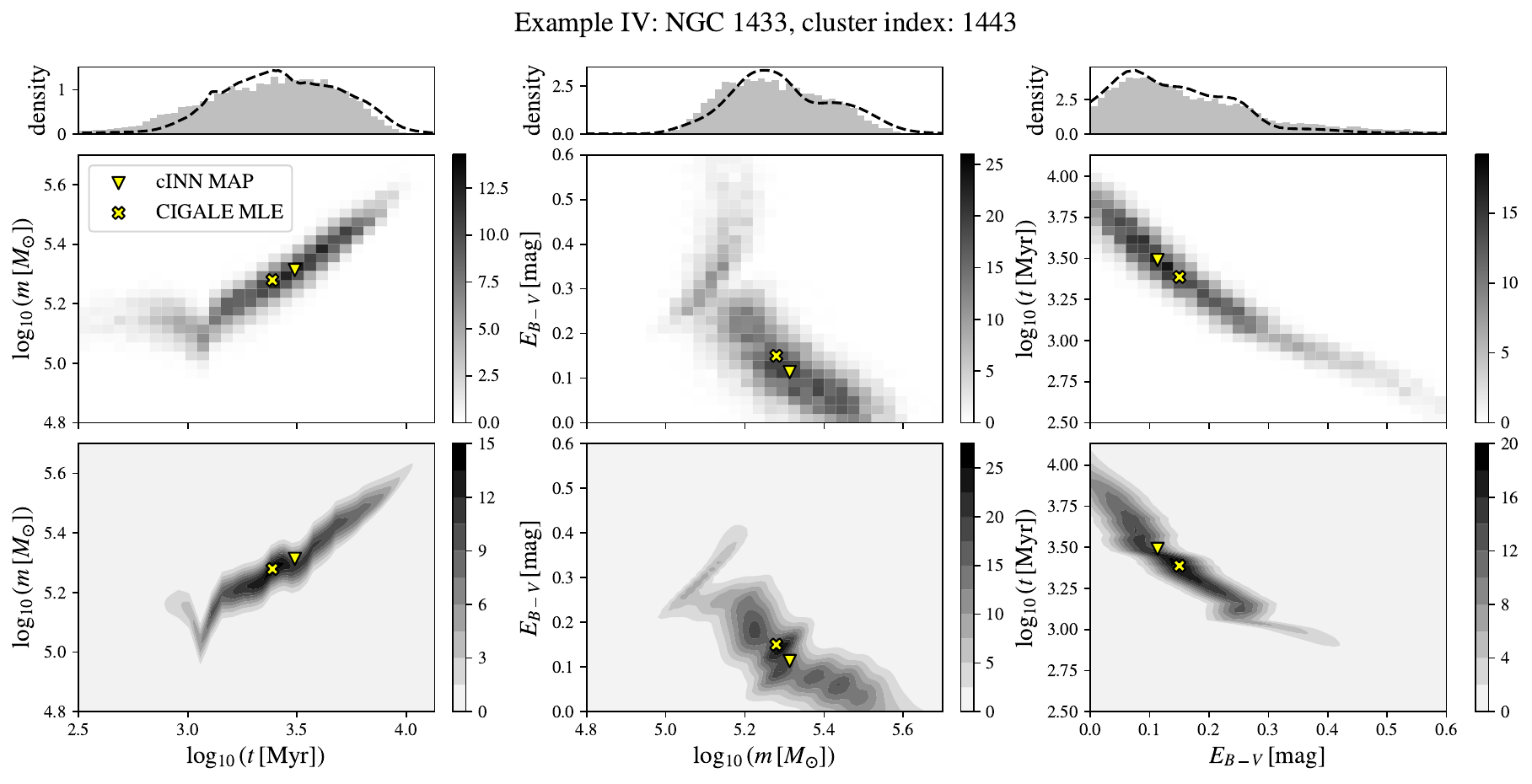}
    \caption*{Continued.}
    \label{results:fig:posterior_examples_2d_continued}
\end{figure*}

Looking at Example I (very young), we see that the posterior distribution shows noticeable artifacts in the age versus mass histogram. These artifacts are a consequence of the fact that our \texttt{CIGALE} version was limited to a discrete age grid with a resolution of 1 Myr. Since the cINN is trained on the \texttt{CIGALE} forward model, it inherits the 1 Myr discretization, which results in grid artifacts. In log-space, this becomes noticeable in the low-age regime, where the logarithm strongly stretches the parameter space.

Example II (young) shows a clearly multimodal distribution. Notice how a difference in mode selection leads to a large discrepancy between our MAP estimate and the PHANGS estimate. Example III (intermediate age) shows a similar multi-modality. However, here, the high-age mode is much more pronounced than the low-age mode. Consequently, the two estimates coincide well. A comprehensive comparison of cINN MAP estimates and PHANGS MLEs is given in Sect. \ref{sec:eval_obs:MAP_MLE}.

Finally, Example IV (old) shows a typical high-age posterior distribution that we included for completeness. We note, however, that the example posteriors presented in this paper cover only a small fraction of the diversity of distributions that we encounter.

\subsubsection{Recovering PHANGS estimates}\label{sec:eval_obs:MAP_MLE}
Since our SED modeling setup is identical to \citet{2021MNRAS.502.1366T} and our prior is uniform, the cINN MAP estimates should correspond to the PHANGS MLEs (see Eq. \ref{eq:equivalence_map_ml}). It therefore makes sense to investigate these estimates for possible deviations.

To find the MAP values of the joint posterior distributions, we used the sampling capabilities of the cINN to generate $10^4$ samples per posterior. We found that in some cases the result can strongly depend on the choice of MAP estimator. For the final evaluation, we settled on the mean shift estimator \citep{1055330}. The related algorithm is mode-seeking in nature and can be used to find additional (non-MAP) modes of the posterior. For every posterior, we determined up to three modes of the multidimensional joint distribution.

\begin{figure*}
    \includegraphics[width=\textwidth]{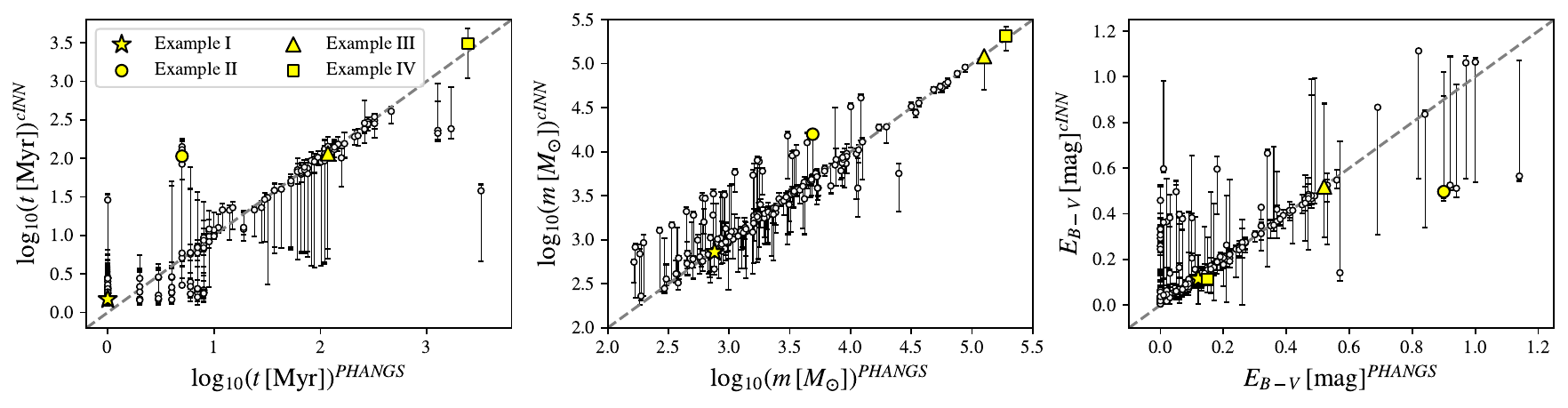}
    \includegraphics[width=\textwidth]{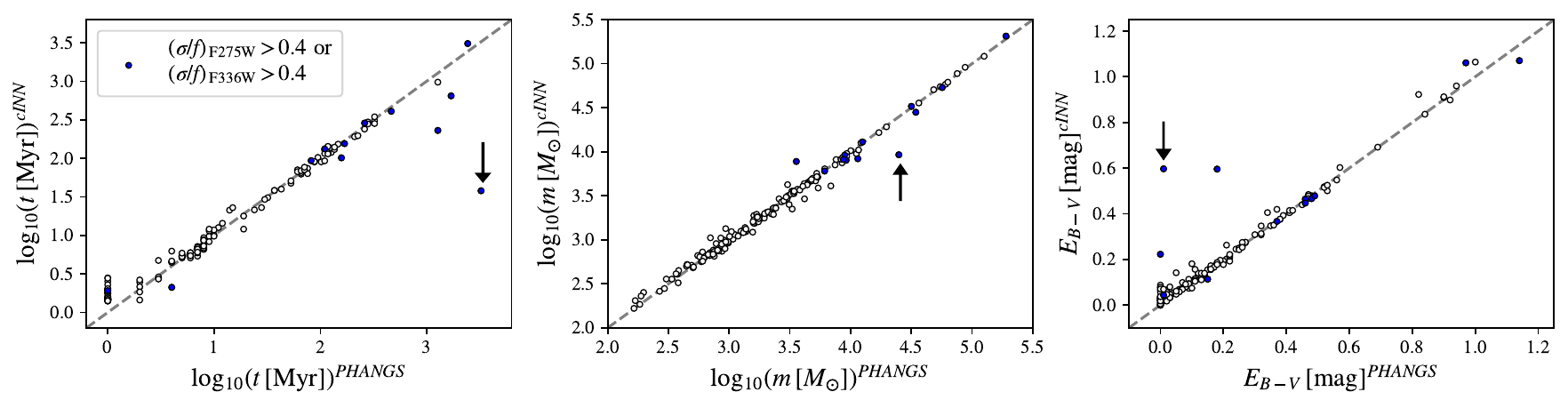}
    \caption{Comparison of cINN posterior modes and PHANGS MLEs for all (filtered) NGC~1433 clusters. Top panels: cINN MAP modes as determined by the mean-shift estimator. Error bars show the 16\%- and 84\%-quantiles of the marginalized posteriors. The yellow markers indicate the positions of the four example clusters. Bottom panels: Comparison with the posterior modes that were closest to the PHANGS MLE (for every posterior, a maximum of three modes were determined). For better visibility, we omitted the error bars. The blue dots mark clusters that have relative flux uncertainties larger than 40\% in at least one UV band, as discussed in Sect. \ref{sec:outliers}. The arrow shows the position of the largest outlier. Comparisons for all galaxies are provided in Appendix \ref{supplementary_plots}.}
    \label{results:fig:cinn_vs_cigale_ngc1433}
\end{figure*}

Figure~\ref{results:fig:cinn_vs_cigale_ngc1433} (top) shows a direct comparison between cINN MAP estimates and PHANGS MLEs for NGC~1433 (Fig.~\ref{results:fig:cinn_vs_cigale} shows the same comparison for all galaxies).
We find that all galaxies show similar patterns of deviation. Interestingly, the age plots show a sword-like shape. \citet{2021MNRAS.502.1366T} find an analogous sword-like pattern when they investigate how small errors in photometry affect age, mass, and reddening estimates on a sample of NGC~3351 clusters \citep[see Sect. 3.2 and Fig. 2 of][]{2021MNRAS.502.1366T}. They report that these deviations are mostly a result of bimodalities in the underlying probability distribution functions. Likewise, we find that the large deviations around 10 Myr are mostly a result of multimodal posterior distributions. To further emphasize that the residuals are mostly a result of differing mode selection, we compared the PHANGS estimates with the peaks of the additional modes that occur in the posteriors. Figure~\ref{results:fig:cinn_vs_cigale_ngc1433} (bottom) plots the posterior modes that are closest to the PHANGS estimates against the PHANGS estimates, resulting in drastically reduced residuals (comparisons for all galaxies can be found in Fig.~\ref{results:fig:best_mode_vs_cigale}). Generally, we find that large residuals tend to be associated with broader posterior distributions (see Fig.~\ref{results:fig:deviation_vs_width}). This again illustrates that deviations between our MAP estimates and the PHANGS MLEs occur mainly for observations that put fewer constraints on the parameters. In Table \ref{table:cINN_vs_PHANGS} we summarize inlier statistics that quantify how often the PHANGS estimates are contained within the cINN marginal posteriors and, conversely, how often the cINN point estimates are contained in the $1\sigma$-regions of the PHANGS estimates.
\begin{table}
\centering
\def\arraystretch{1.5}
\caption{Inlier percentages for cINN and PHANGS estimates.}
\begin{tabular}{c |c c c } 
 & $\log_{10}(t)$ & $\log_{10}(m)$ & $E_{B-V}$ \\\hline
 $Q_{0.16} < \theta <  Q_{0.84}$ & $71.0\%$ & $61.1\%$ & $75.4\%$ \\
 $\theta - 1\sigma <  \Bar{\theta}  < \theta + 1\sigma$ & $65.4\%$ & $52.0\%$ & $68.4\%$  \\
  $\theta - 1\sigma < Q_{\frac{1}{2}} < \theta + 1\sigma$ & $73.3\%$ & $59.8\%$ & $77.8\%$ \\
  $\theta - 1\sigma < \theta_{MAP} < \theta + 1\sigma$ & $76.3\%$ & $64.3\%$ & $80.5\%$  
\end{tabular}
\tablefoot{The top row shows how often the PHANGS estimates ($\theta$) are enclosed by the 16\%- to 84\%-quantiles of the cINN posterior. Rows two, three and four show how often the cINN posterior mean ($\Bar{\theta}$), median ($Q_{\frac{1}{2}}$), and MAP estimates ($\theta_{MAP}$) are contained in the $\pm$ 1$\sigma$-range of the PHANGS estimates.}
\label{table:cINN_vs_PHANGS}
\end{table}

\begin{figure*}
    \centering
    \includegraphics[width=\textwidth]{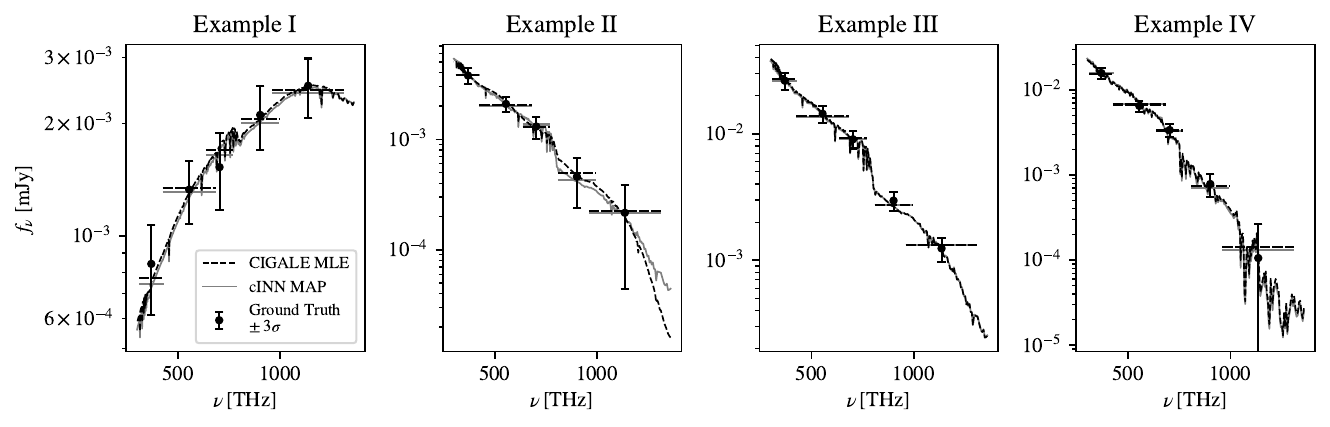}
    \caption{Simulated SEDs for the example clusters (cINN MAP and PHANGS MLE). The horizontal dashed lines indicate the broadband fluxes for the five HST filters. The black dots and error bars show the ground truth photometry and $\pm$ 3$\sigma$ range as found in the PHANGS catalog.}
    \label{results:fig:re-simulation_example}
\end{figure*}

\begin{figure}
    \centering
    \includegraphics[width=0.5\textwidth]{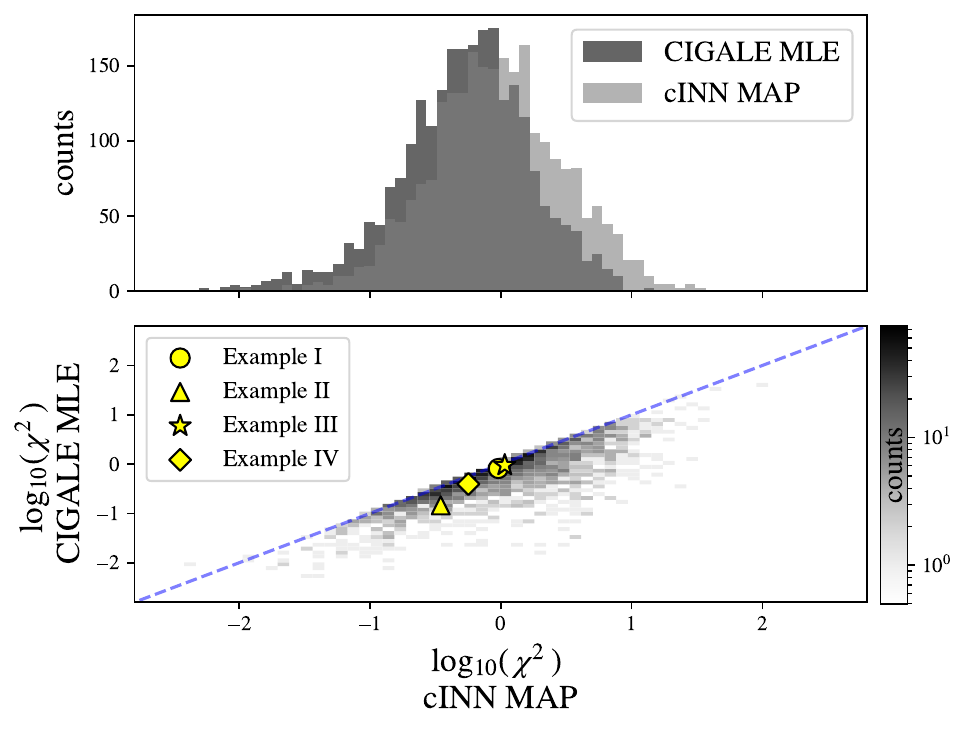}
    \caption{Comparison of reduced $\chi^2$ deviations for the cINN MAP estimates and the PHANGS MLEs. Top panel: Histograms of the $\log_{10}(\chi^2)$ distributions. Bottom panel: PHANGS MLE deviations versus the cINN MAP deviations. The differences are somewhat exaggerated by the logarithmic color scale. Yellow markers indicate the $\log_{10}(\chi^2)$ values of the example clusters.}
    \label{results:fig:chi2}
\end{figure}

To test the quality of the generated posterior samples, we used \texttt{CIGALE} to re-simulate the corresponding photometry for the retrieved cluster parameters. Ideally, the re-simulated photometry should be contained within the error range of the initial observations. Figure~\ref{results:fig:re-simulation_example} shows the re-simulated SEDs of the four example clusters introduced in Sect.~\ref{sec:eval_obs:MAP_MLE}.
Visually, it appears that the re-simulated SEDs of examples I, II, III, and IV are well within the $ \pm 3\sigma$-range of the initial observation for both cINN and PHANGS estimates. To quantify the residuals, it is useful to calculate the (reduced) $\chi^2$-deviations between re-simulated photometry and ground truth observations:
\begin{equation}
    \chi^2 = \frac{1}{\nu} \cdot \sum_i \left(\frac{f_i^{\mathrm{\:resim}} - f_i}{\sigma_i}\right)^2,
\end{equation}
where $\nu$ corresponds to the degrees of freedom.
Figure~\ref{results:fig:chi2} shows the $\chi^2$-deviations of re-simulated MAP estimates for the entire PHANGS catalog.
Primarily, these plots indicate that the cINN MAP estimates exhibit larger $\chi^2$-deviations than the PHANGS MLEs. This is expected since the PHANGS estimates are based on a minimum $\chi^2$ fit (and are hence optimized to minimize the $\chi^2$-deviation between observation and re-simulation). The cINN, on the other hand, finds these estimates only in a very indirect way by modeling the entire probability density function (by minimizing the KL-divergence between the model posterior and target posterior) and determining the MAP estimates as a byproduct. Nevertheless, we find that these values are acceptable.

\subsubsection{Outliers and the effect of a log-normal noise model}\label{sec:outliers}
\begin{figure}
    \centering
    \includegraphics[width=\linewidth]{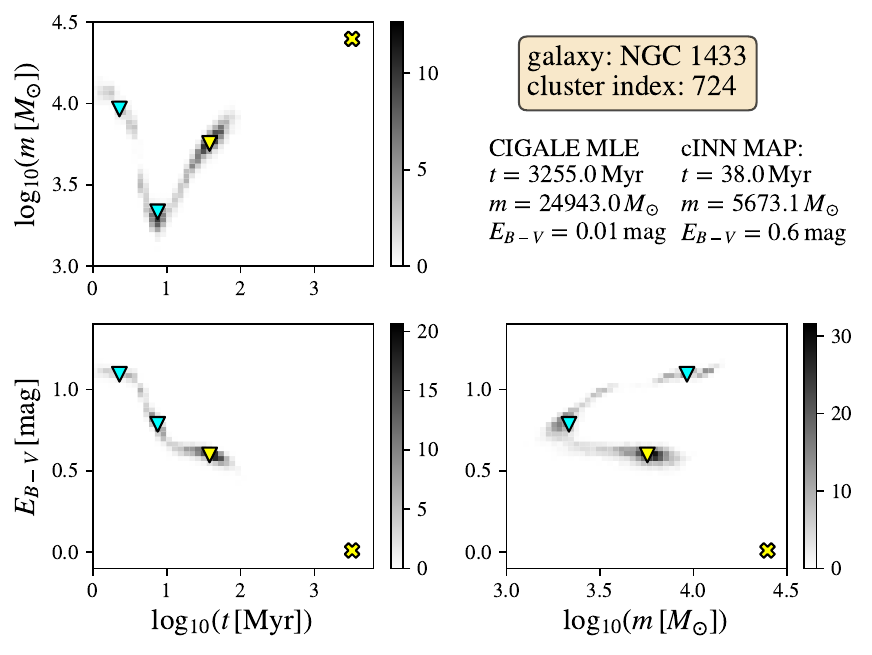}
    \caption{Posterior distribution generated by the cINN for the outlier cluster (marked with an arrow) in Fig. \ref{results:fig:cinn_vs_cigale_ngc1433}. Triangles show positions of the posterior modes; the yellow triangle denotes the MAP value and cyan triangles denote the second and third mode. The yellow cross shows the position of the \texttt{CIGALE} MLE. The numerical values of the parameter estimates are shown in the top right.}
    \label{results:fig:outlier_posterior}
\end{figure}

Figure \ref{results:fig:cinn_vs_cigale_ngc1433} illustrates that a small number of clusters show large deviations even when we compare the closest mode of the cINN posterior with the \texttt{CIGALE} MLE. As it turns out, this is mostly an artifact of the differing noise models. As an example, take the cluster with ID 724 in NGC 1433. Figure \ref{results:fig:outlier_posterior} shows the posterior that the cINN generates for this cluster. As we can see, the \texttt{CIGALE} MLE is fully outside the support of the posterior. We mark the position of the cluster in Fig. \ref{results:fig:cinn_vs_cigale_ngc1433} with an arrow. Clearly, this is the largest outlier in NGC 1433. At the same time, it is the only cluster in NGC 1433 that has relative flux uncertainties above 40\% in both UV bands. In fact, we find that all the other major outliers are also associated with large UV band uncertainties. In Fig. \ref{results:fig:cinn_vs_cigale_ngc1433} we mark every cluster in blue that has a relative flux uncertainty larger than 40\% in at least one UV band. Figure \ref{results:fig:flux_uncertainty_statistics} (bottom row) makes it apparent why the UV bands are of particular importance here. They tend to have much larger relative flux uncertainties and are therefore the only bands for which the log-normal noise model deviates strongly from the normal noise model. This shows that for some cases, the noise model assumptions can have a large impact on the cluster parameter estimates.

\section{Discussion}\label{sec:discussion}
The aim of this paper is to explore the network's behavior on a simplified cluster inference problem and to establish that the cINN can produce accurate estimates of cluster age, mass, and reddening from broadband photometric observations. Our analysis shows that this is the case, both for synthetic and real-world photometry. Furthermore, the ability to model the full joint posterior distribution enables accurate uncertainty estimation and can be used as a basis for further statistical analysis (e.g., via the calculation of correlations, moments, quantiles, or entropy related quantities).

In the following we further justify our normalizing flow approach by detailing some computational and statistical aspects of this inference method. 

\subsection{Sampling and density estimation}
When modeling the posterior distribution, we would like to perform two similar but distinct tasks. On the one hand, we want the ability to sample from the underlying probability distribution $\boldsymbol{\theta}_1, \dots, \boldsymbol{\theta}_n \sim p(\boldsymbol{\theta} \mid \mathbf{x})$. This enables, for example, the efficient computation of the moments of the posterior via $m_k = \mathbb{E} \left[ \boldsymbol{\Theta}^k \right] \approx \frac{1}{n} \sum_i \boldsymbol{\theta}_i^k$. On the other hand, it is often interesting to assess the density of the posterior. For instance, if we want to estimate the entropy (and related quantities like KL-divergence or Mutual Information) of the posterior distribution, we need a model for the density of the posterior. In fact, combining sampling capabilities with density estimation capabilities can lead to a Monte Carlo-based estimator of the differential entropy: $h(\boldsymbol{\Theta}\mid\mathbf{X} = \mathbf{x}) = \int p(\boldsymbol{\theta}\mid\mathbf{x}) \cdot \log p(\boldsymbol{\theta}\mid\mathbf{x}) \:d\boldsymbol{\theta} \approx \frac{1}{n} \sum_i \log p(\boldsymbol{\theta}_i \mid \mathbf{x})$. In general, sampling and density estimation provide different advantages and can enable the application of different methods.

Normalizing flows differentiate themselves from other methods (e.g., variational autoencoders or generative adversarial networks) by providing the capability for both efficient sampling and density estimation. The former can be achieved by first sampling the latent variable distribution and then applying the forward pass of the network, i.e., $f_{\boldsymbol{\phi}}(\mathbf{z})$ with $\mathbf{z} \sim p_{\mathbf{Z}}$. Conversely, for density estimation, we used the Jacobi determinant in conjunction with the backward pass: $p_{\boldsymbol{\Theta}}(\boldsymbol{\theta}) \approx p_{\mathbf{Z}}(f_{\boldsymbol{\phi}}^{-1}(\boldsymbol{\theta}))/|\det J_{f_{\boldsymbol{\phi}}} (f_{\boldsymbol{\phi}}^{-1}(\boldsymbol{\theta}))|$. Note that different normalizing flow architectures provide different efficiencies when it comes to forward and backward pass. For example, masked autoregressive flows provide an efficient backward pass and thereby efficient density estimation, but have a computationally expensive forward pass, especially for high-dimensional data \citep{NIPS2017_6c1da886}. Coupling-based architectures (like the cINN) on the other hand, have a symmetric computational cost for forward and backward pass. This makes them ideal for situations in which both sampling and density estimation are of interest. This, however, comes at the cost of a lower expressibility per flow layer \citep[][Sect. 3.1.2]{JMLR:v22:19-1028}.
\subsection{Comparison to Markov chain Monte Carlo}
A popular inference method that provides sampling capabilities is the Metropolis–Hastings algorithm, which constitutes an MCMC method. In fact, \citet{2011ApJ...740...22S} have successfully used the Metropolis-Hastings algorithm to infer galaxy properties with \texttt{CIGALE} \citep[using an older version of the code as described in][]{2009A&A...507.1793N}. They emphasize that the computational cost of a grid-based approach scales exponentially with the number of inferred parameters and is therefore costly compared to the Metropolis-Hastings approach.

To our knowledge, there are no formal mathematical results pertaining to the scaling of the computational cost of normalizing flow methods when applied to high-dimensional data. However, numerous empirical results show that popular flow architectures can model high-dimensional distributions. For example, \citet[Sect. 4]{9089305} provide a summary of log-likelihood scores for common flow architectures on high-dimensional ($D = 6, 8, 21, 43, 63$) tabular datasets and high-dimensional ($D = 784, 3072, 12288$) image datasets.

In this work, the number of inference parameters was low ($D=3$). However, in future works we may increase this number (e.g., by additionally inferring metallicity and other insightful parameters). The scaling of the computational cost is therefore of practical interest. The considerations above show that it is unclear how the normalizing flow approach compares to classical MCMC approaches in this regard.

Another crucial aspect relates to the tractability of the likelihood function. The classical Metropolis-Hastings algorithm (as applied to Bayesian inference) generates samples of the posterior by drawing from a Markov process with transition probability 
\begin{equation}
    A(\boldsymbol{\theta}_i, \boldsymbol{\theta}') = \min \left(1, \frac{p(\mathbf{x}\mid \boldsymbol{\theta}') \cdot p(\boldsymbol{\theta}') \cdot q(\boldsymbol{\theta}_i \mid \boldsymbol{\theta}')}{p(\mathbf{x} \mid \boldsymbol{\theta}_i) \cdot p(\boldsymbol{\theta}_i) \cdot q(\boldsymbol{\theta}' \mid\boldsymbol{\theta}_i)}\right),
\end{equation}
where $q(\boldsymbol{\theta}_i \mid \boldsymbol{\theta}')$ denotes the proposal distribution of the current state $\boldsymbol{\theta}_i$ and proposed state $\boldsymbol{\theta}'$. Calculating the transition probability requires the tractability of the likelihood $p(\mathbf{x} \mid \boldsymbol{\theta})$. As discussed in Sect. \ref{sec:introdution}, we believe that the inclusion of additional nuisance parameters can lead to an expensive or intractable likelihood function. This can make the classical Metropolis-Hastings algorithm (and other likelihood-based MCMC methods) infeasible. Normalizing flow methods are trained in a likelihood-free manner and therefore require only a data generation process (i.e., a simulation). In particular, they do not require the explicit calculation of the likelihood. This makes them ideal tools for likelihood-free inference.

In Sect. \ref{sec:introdution} we note that normalizing flows constitute an amortized approach to Bayesian inference. This means that a high initial computational cost (i.e., training) is compensated by a smaller computational cost during inference. Therefore, amortization becomes useful when the number of inferences and computational cost per simulation become sufficiently large. MCMC methods, on the other hand, require re-simulations for every inference. This can make normalizing flow approaches significantly cheaper for scenarios with many inferences \citep[see, e.g.,][]{2022ApJ...938...11H}.

Finally, we should mention that adaptations to the classical MCMC method have led to amortized and likelihood-free MCMC approaches. For example, surrogate models can amortize the simulation cost \citep[e.g.,][]{2022PSJ.....3...91H} and ``amortized ratio estimators'' can be used to extend MCMC methods to likelihood-free scenarios \citep{pmlr-v119-hermans20a}. The field of MCMC methods is large, and a discussion of these techniques would exceed the scope of this paper.

\section{Summary}\label{sec:summary}
We used the SED modeling code \texttt{CIGALE} to generate a training set of $5 \times 10^6$ sets of cluster parameters (age, mass, and reddening) and photometric fluxes (F275W, F336W, F438W, F555W, and F814W). This dataset was used to train a cINN with the goal of modeling the posterior distribution of the cluster parameters conditioned on the observed photometry.

After the training phase of the cINN, we performed various tests to ascertain the performance of the network. In our SBC analysis on the synthetic dataset, we find a reasonably uniform rank distribution over six selected statistics, with some deviations from uniformity that become apparent in the ECDF difference plots. In addition, we performed PPCs on the real-world photometry. Here, we find that the mean of the posterior predictive distribution aligns well with the initial photometry and that the corresponding rank distributions behave as expected. When we compare the cINN MAP estimates with the PHANGS MLEs, we find that deviations with respect to the \citet{2021MNRAS.502.1366T} estimates are mostly a result of differing mode selection, except for some outlier cases that are caused by different noise model assumptions. To assess the quality of our estimates, we re-simulated our MAP estimates and found that the cINN produced larger $\chi^2$ deviations compared to the PHANGS MLEs.

Overall, we find that the cINN generates decent approximations of the cluster posterior distributions. However, we must keep an important caveat in mind. If the goal is solely to generate parameter estimates with a model that uses a fully realized IMF and no additional model nuisance parameters (i.e., a case in which the grid-based approach is very efficient), then we cannot expect to gain much from the normalizing flow approach. In fact, this work shows that the cINN produces consistently worse MLEs compared to traditional grid-based fitting. However, the cINN will allow us to model posterior distributions in scenarios that are intractable to classic grid-based approaches. Here, we are referring in particular to models that incorporate multiple nuisance parameters and add additional inference parameters, which will be the focus of our follow-up paper

\begin{acknowledgements}
DW, VFK, and RSK acknowledge financial support from the European Research Council via the ERC Synergy Grant ``ECOGAL'' (project ID 855130),  from the German Excellence Strategy via the Heidelberg Cluster of Excellence (EXC 2181 - 390900948) ``STRUCTURES'', and from the German Ministry for Economic Affairs and Climate Action in project ``MAINN'' (funding ID 50OO2206). The team at Heidelberg University is also  grateful for computing resources provided by the Ministry of Science, Research and the Arts (MWK) of the State of Baden-W\"{u}rttemberg through bwHPC and the German Science Foundation (DFG) through grants INST 35/1134-1 FUGG and 35/1597-1 FUGG, and also for data storage at SDS@hd funded through grants INST 35/1314-1 FUGG and INST 35/1503-1 FUGG.
MB acknowledges support by the ANID BASAL project FB210003. This work was supported by the French government through the France 2030 investment plan managed by the National Research Agency (ANR), as part of the Initiative of Excellence of Université Côte d’Azur under reference No. ANR-15-IDEX-01. This research was funded, in whole or in part, by the French National Research Agency (ANR), grant ANR-24-CE92-0044 (project STARCLUSTERS).  We thank the German Science Foundation DFG for financial support in the project STARCLUSTERS (funding ID KL 1358/22-1 and SCHI 536/13-1).
KG is supported by the Australian Research Council through the Discovery Early Career Researcher Award (DECRA) Fellowship (project number DE220100766) funded by the Australian Government.
\end{acknowledgements}

\bibliographystyle{aa}
\bibliography{bibliography}

\onecolumn
\appendix
\section{Network parameters}\label{network_parameters}
{\rowcolors{2}{gray!15}{white}
\def\arraystretch{1.5}
\begin{table}[h]
\begin{center}
\caption{Network architecture and training parameters.}
\begin{tabular}{ c|m{3cm}|c|c|c } 
 \textbf{parameter} & \textbf{explanation} & \textbf{fixed?} & \textbf{search range} & \textbf{final value(s)} \\ 
 \hline
 number of coupling blocks & The total number of invertible conditional coupling blocks (see Sect. \ref{sec:cINN recap}).  & no & $[4,\: 21]$ & 12\\
 internal network depth & The number of hidden layers in the coupling block subnetworks. & yes & - &  2\\
 internal network width & Number of neurons per layer in coupling block subnetworks. & no & $[128,\: 2048]$ &  256\\
 activation function & Activation function of subnetworks. & yes & - &  ReLU\\
 initial learning rate & Step size of optimizer at first epoch. & no & $[10^{-5},\: 10^{-2}]$ &  0.001876\\
 learning rate decay & Factor by which learning rate decays after a fixed number of meta-epochs.& no & $[0.1,\: 0.9]$ &  0.6637\\
 $\beta_1, \: \beta_2$ & Parameters of Adam optimizer \citep[see][]{2014arXiv1412.6980K}. & no & $(0.8,\: 0.8)$ or $(0.9,\: 0.999)$ &  $(0.9,\: 0.999)$\\
 batch size & Number of (random) training samples in a single optimization step. & no & $[64,\: 1024]$ &  512\\
 $\eta$ & Regularization strength as described in Sect. \ref{sec:cINN recap}.  & no & $[10^{-7}, \: 10^{-4}]$ &  $1.353 \times 10^{-6}$\\
\end{tabular}
\tablefoot{The third column indicates whether a parameter has been subject to the hyperparameter search. In that case, a search range is specified.}
\end{center}
\end{table}

\newpage
\section{Supplementary plots}\label{supplementary_plots}
\begin{figure*}[h]
    \centering
    \includegraphics[width=\textwidth]{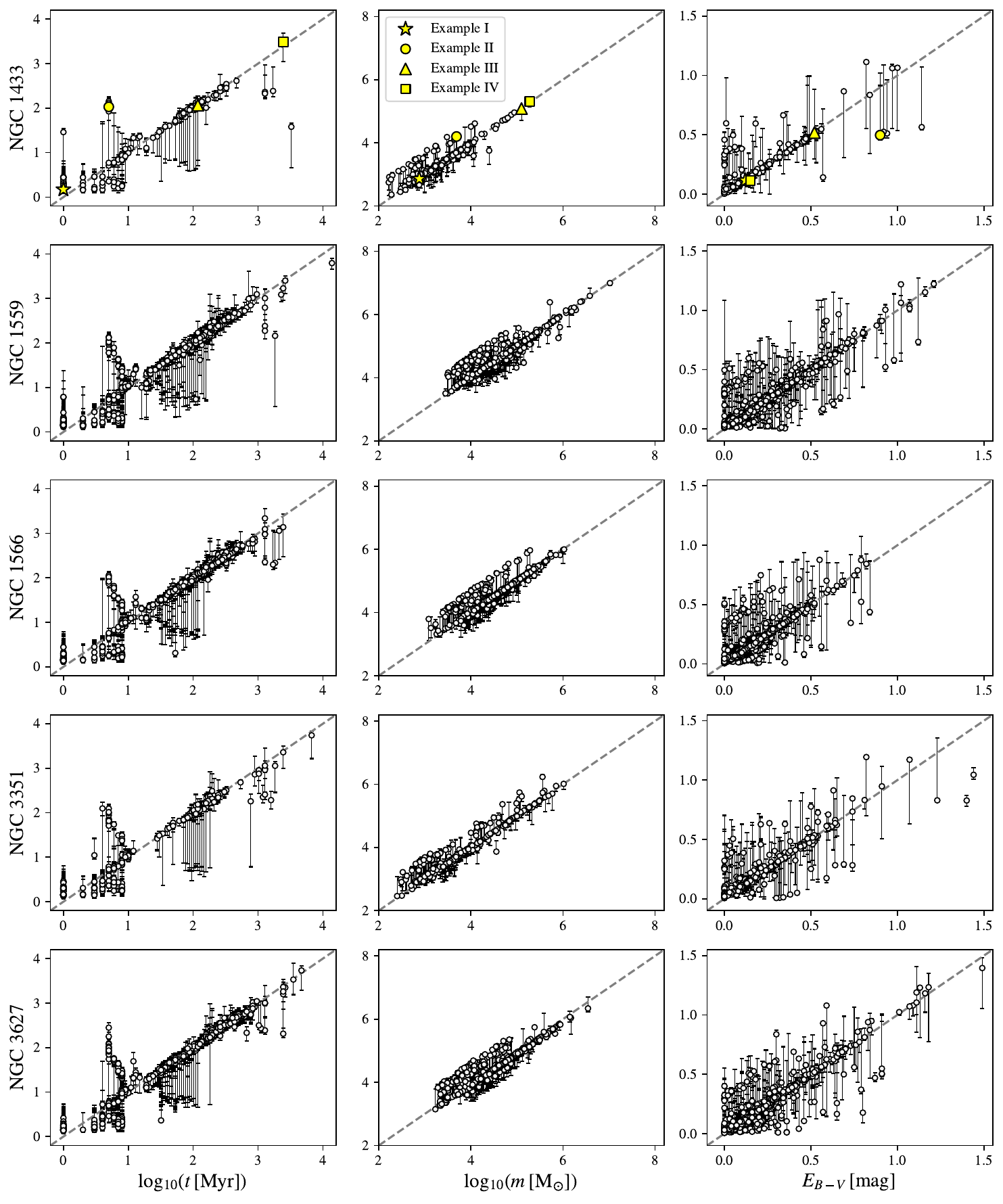}
    \caption{Same as Fig. \ref{results:fig:cinn_vs_cigale_ngc1433} (top) but including all galaxies from the PHANGS catalog.}
    \label{results:fig:cinn_vs_cigale}
\end{figure*}

\begin{figure*}[h]
    \centering
    \includegraphics[width=\textwidth]{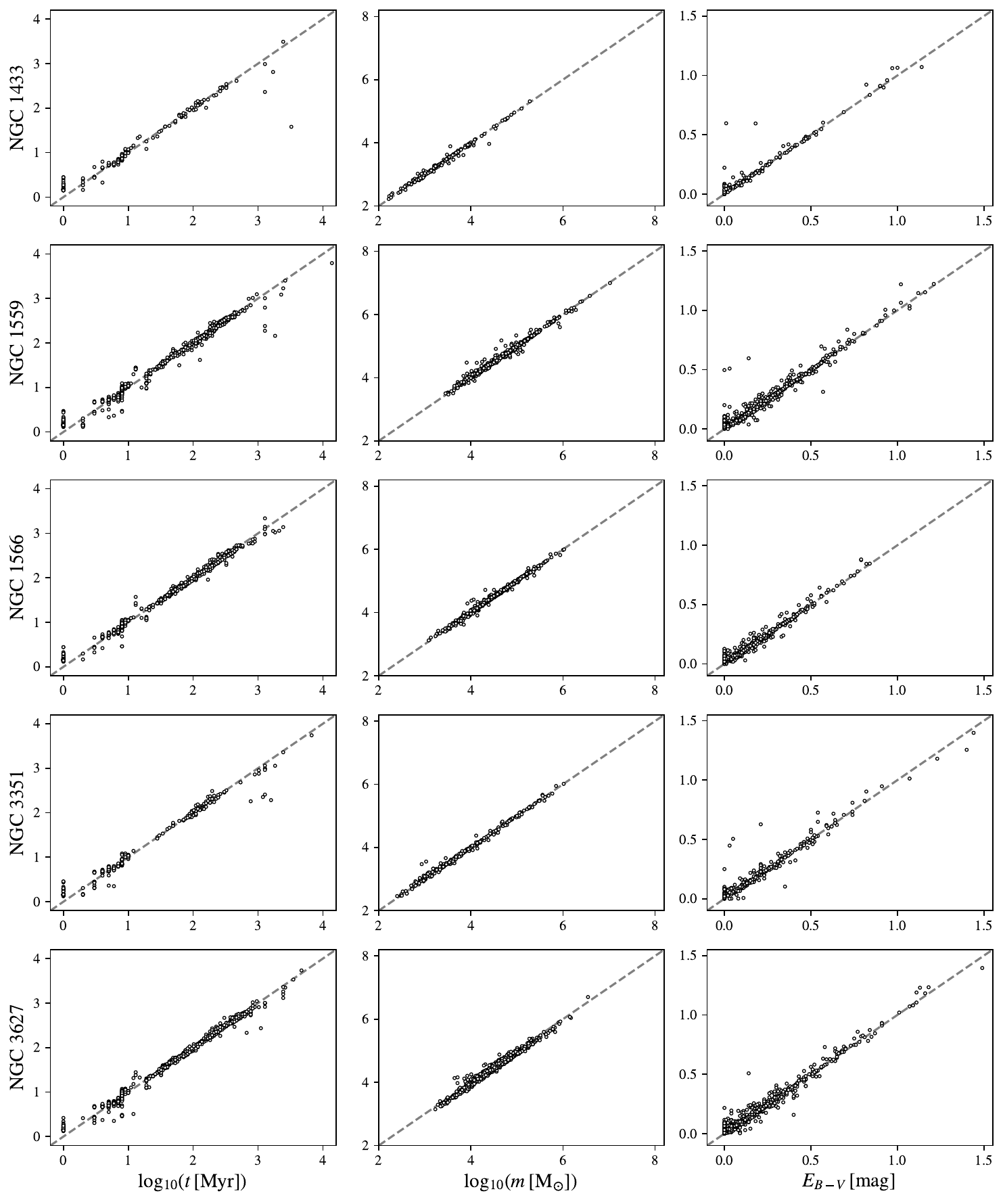}
    \caption{Same as Fig. \ref{results:fig:cinn_vs_cigale_ngc1433} (bottom) but including all galaxies from the PHANGS catalog.}
    \label{results:fig:best_mode_vs_cigale}
\end{figure*}

\begin{figure*}[h]
    \centering
    \includegraphics[width=\textwidth]{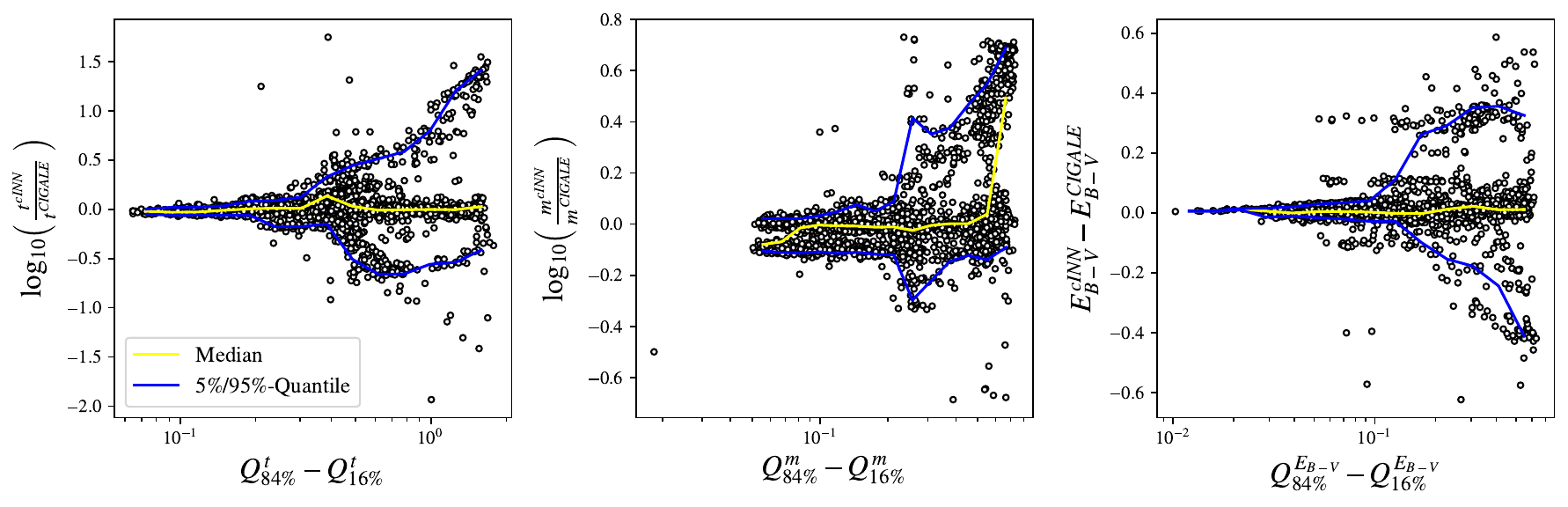}
    \caption{Deviations between cINN MAP estimates and PHANGS MLEs plotted against the posterior width (as quantified by the distance between 16\% and 84\% quantile) for the entire cluster catalog. The yellow and blue lines indicate the median, 5\% and 95\% quantiles of the deviations.}
    \label{results:fig:deviation_vs_width}
\end{figure*}

\end{document}